\documentclass[sigconf, nonacm]{acmart}

\usepackage{booktabs} 
\usepackage[ruled,vlined]{algorithm2e}

\usepackage{enumitem}

\usepackage[ruled]{algorithm2e} 

\SetAlFnt{\small}
\SetAlCapFnt{\small}
\SetAlCapNameFnt{\small}
\SetAlCapHSkip{0pt}

\usepackage{caption}
\usepackage{subcaption}
\usepackage{bm}
\usepackage{multirow}
\usepackage{colortbl}
\usepackage{xcolor}
\usepackage{makecell}
\LinesNumbered

\newcommand{\norm}[1]{\|#1\|}

\definecolor{mygray}{rgb}{0.2 .2 .5}

\SetCommentSty{mycommfont}

\DeclareMathOperator*{\argmin}{\arg\!\min}
\newcommand\Bstrut{\rule[-0.9ex]{0pt}{0pt}}   

\begin{document}
\title{Generalized Spectral Coarsening}

\author{Alexandros D. Keros}
\orcid{}
\affiliation{%
  \institution{The University of Edinburgh}
  \streetaddress{10 Crichton Street}
  \city{Edinburgh}
  \state{Scotland}
  \postcode{EH8 9AB}
  \country{UK}}
\email{a.d.keros@sms.ed.ac.uk}

\author{Kartic Subr}
\orcid{}
\affiliation{%
  \institution{The University of Edinburgh}
  \streetaddress{10 Crichton Street}
  \city{Edinburgh}
  \state{Scotland}
  \postcode{EH8 9AB}
  \country{UK}}
\email{k.subr@ed.ac.uk}

\begin{teaserfigure}
	\begin{center}		
	\includegraphics[width=.98\linewidth]{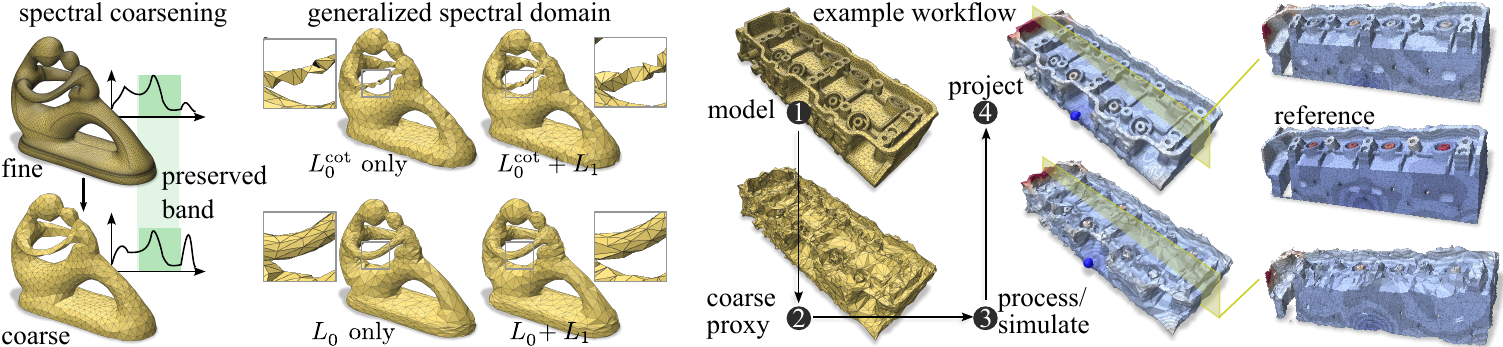}
	\caption{\label{fig:teaser} We present a method to simplify geometry while preserving targeted bandwidths in a customizable spectral domain. The \emph{generalized spectral domain} is an input to our algorithm and can be defined as a combination of multiple discrete Laplacian operators. Our method includes a simple ``lifting'' operator to enable the workflow shown on the right. The example shows the calculation of biharmonic distances over a tetrahedral mesh of an engine part with 20 small \emph{holes} that are preserved.}
	\end{center}
\end{teaserfigure}


\begin{abstract}

Many computational algorithms applied to geometry operate on discrete representations of shape. It is sometimes necessary to first simplify, or \emph{coarsen}, representations found in modern datasets for practicable or expedited processing. The utility of a coarsening algorithm depends on both, the choice of representation as well as the specific processing algorithm or operator. e.g. simulation using the Finite Element Method, calculating Betti numbers, etc. We propose a novel method that can coarsen triangle meshes, tetrahedral meshes and simplicial complexes. Our method allows controllable preservation of salient features from the high-resolution geometry and can therefore be customized to different applications. 

Salient properties are typically captured by local shape descriptors via linear differential operators -- variants of \emph{Laplacians}. Eigenvectors of their discretized matrices yield a useful \emph{spectral domain} for geometry processing (akin to the famous Fourier spectrum which uses eigenfunctions of the derivative operator).  Existing methods for spectrum-preserving coarsening use zero-dimensional discretizations of Laplacian operators (defined on vertices). We propose a generalized spectral coarsening method that considers multiple Laplacian operators defined in \emph{different dimensionalities} in tandem. Our simple algorithm greedily decides the order of contractions of simplices based on a quality function per simplex. The quality function quantifies the error due to removal of that simplex on a chosen band within the spectrum of the coarsened geometry. 

\end{abstract}

%

%
%


\maketitle

\section{Introduction}\label{sec:intro}

Discrete representations of shape are ubiquitous across computer graphics applications. Meshes, a common choice for computer graphics applications, are specific instances of general abstractions called \emph{simplicial complexes}. While the vertices of a mesh are commonly embedded (have explicit coordinates) in 2D or 3D, simplicial complexes capture abstract relationships between nodes -- as extensions of graphs by including 3-ary (triangles), 4-ary (tetrahedra) and higher dimensional-relationships. We develop an algorithm to coarsen simplicial complexes of arbitrary dimensionality.    

The choice of discretization has a profound impact on downstream applications that operate on the geometry, both in terms of efficiency as well as numerical stability.  \emph{Simplification schemes} are used to reduce the number of discrete elements while preserving quality. Quality can defined either in terms of aesthetic appeal (direct rendering, visualization, etc.) or functionally (finite element simulation, topological data analysis, geometry processing algorithms, etc.). 
We focus on the latter and propose a coarsening algorithm that can be tailored to different applications via a versatile definition of functionally salient features .

We resort to classical spectral theory to formalize the definition of qualities of the input representation which should be preserved while coarsening. Just as the famous Fourier spectrum is obtained via projection onto the eigenfunctions of the univariate derivative operator, our spectral domain of choice is defined via a projection onto a Laplacian operator.
Unlike the univariate  derivative operator, a variety of definitions exist for Laplacians. For discrete Laplacians, these usually arise from the elements of choice (vertices, edges, triangles, etc.) and their associated weights. 
The choice of the variant of Laplacian considered impacts the utility of the coarsened mesh in applications. For example, 
simplification of the domain of a physics simulation requires the preservation of  the spectra of \emph{both 0- and 1-dimensional Laplacians}~\cite{dewittFluidSimulationUsing2012a}.
The input to our algorithm includes a list of Laplacians and associated spectral bands -- each band is a subspace in the spectrum corresponding to a Laplacian. Its output is a coarsened representation that preserves the spectral profile in the specified bands.

%

To summarize, we propose a coarsening algorithm for simplicial complexes that can preserve  spectral bands of different Laplacians, across different dimensionalities of simplices.  Our simple algorithm operates by first evaluating a quality function per simplex, which quantifies the error introduced by contracting (eliminating) that simplex towards the specified spectral band(s) to be preserved. Then, we greedily perform contractions iteratively by choosing candidates with minimum error until the target coarsening level is reached. Thus, our coarsening algorithm is agnostic to the specific Laplacian considered. 
Our contributions in this paper are:
\begin{itemize}[topsep=0pt,itemsep=0pt, leftmargin=*]
	\item a novel coarsening operator that is Laplacian-independent;
	\item a coarsening operator that simultaneously preserves spectral bands associated with multiple Laplacians;
	\item an algorithm for band-pass filtering of simplicial complexes.
\end{itemize}
We evaluate our method using a variety of surface (triangular) and volumetric (tetrahedral) meshes, as well as simplicial complexes.

\section{Related Work}\label{sec:relatedwork}

\paragraph{Graphs} The spectrum of a combinatorial graph Laplacian reveals fundamental geometric and algebraic properties of the underlying graph~\cite{Chung1999spectral}. Several works attempt to preserve spectral subspaces, while reducing the size of input graphs~\cite{chen2022graph}. A notable example~\cite{loukas2019graph} proposes an iterative, parallelizable solution to preserve spectral subspaces of graphs by minimizing undesirable projections.

\paragraph{Meshes} Seminal works for coarsening triangle meshes propose localized and iterative operations via edge collapses based on visual criteria~\cite{garland1997surface,ronfard1996full}. Similar methods have also been applied to tetrahedral mesh simplification,  based on volume, quadric-based, and isosurface-preserving criteria~\cite{chopra2002tetfusion, voStreaming2007, chiangProgressive2003a}. Recent methods formulate coarsening as an optimization problem subject to various sparsity conditions~\cite{liu2019spcoarsegeomops}, by detaching the mesh from the operator~\cite{chen2020chordal} and localizing error computation to form a parallelizable strategy~\cite{lescoat2020spectral}. The \emph{cotan Laplacian} is a popular choice, and has also been used, via its  functional maps, to identify correspondences between partial meshes~\cite{rodola2017partial} .

\paragraph{Simplicial complexes and computational topology} 
The \emph{link condition}~\cite{dey1998topology} is a combinatorial criterion ensuring that homology is preserved while performing strong collapses (merging vertices as opposed to edge removal), with extended applications and extensions to persistent homology~\cite{wilkerson2013simplifying,boissonnat2019computing}. Edge collapses in the form of edge removals~\cite{boissonnat2020edge,glisse2022swap} is a state-of-the-art method for simplifying filtered flag simplicial complexes while preserving their (persistent) homology. These methods focus on the dimensionality of the null space (kernel) of the Laplacian. They rarely investigate the general spectral profile of the reduced complex. Notable exceptions~\cite{osting2017ercomplexes, black2021erquantum,hansen2019ersheaves} apply the method of effective resistances~\cite{spielman2011graph} for coarsening complexes, and cellular sheaves, respectively.

\paragraph{Spectral analysis} 
 The Laplacian makes frequent appearances across geometry processing, machine learning and computational topology.
Its specific definitions and flavours vary widely across domains, such as  
discrete exterior calculus~\cite{desbrun2005discrete, crane2013digital},
vector-field processing~\cite{vaxman2016fields,poelke2016boundary,deGoes2016vfp,wardetzky2020discrete,zhao20193dhodge}, 
fluid simulation~\cite{liu2015model}, mesh segmentation and editing~\cite{lai2008fast, khan2020surface, sorkine2004laplacian}, 
topological signal processing~\cite{barbarossa2020topological}, 
random walk representations~\cite{lahav2020meshwalker, schaub2020random}, 
clustering and learning~\cite{Su2022commdet, nascimento2011spectral,ebli2019notion,ebli2020simplicial, keros2022dist2cycle,smirnov2021hodgenet}. Its ability to effectively capture salient geometric, topological, and dynamic information of the object of interest makes its spectrum a versatile basis. 

The de-facto Laplacian operator used in mesh processing is the Laplace-Beltrami operator which is approximated via a discretization called the \emph{cotan Laplacian} defined on vertices (0D)  with vertex and edge weights. A multitude of other Laplacians have been shown to accommodate non-manifold meshes~\cite{sharp2020laplacian}, FEM simulations~\cite{ayoub2020new}, digital surfaces~\cite{caissard2019laplace}, polygonal meshes~\cite{bunge2021diamond} and arbitrary simplicial complexes~\cite{ziegler2022balanced}.

\paragraph{Motivation}
The variety of definitions and properties of Laplacian operators (see Figure~\ref{fig:teaser}) suggests that a unified spectral coarsening algorithm could impact a range of applications. Our Laplacian-agnostic spectral coarsening can be tailored by considering the weightings as special cases of \emph{Hodge Laplacians}.
We achieve this by adapting the cost function proposed for graph theory~\shortcite{loukas2019graph} to simplicial complexes and incorporating it within a standard mesh-coarsening framework~\cite{garland1997surface}.

\section{Background}\label{sec:background}

\subsection{Simplicial Complexes \& Meshes}
A \emph{simplicial complex} $K$ is constructed by considering appropriate subsets of a finite set $V$ of \emph{vertices}. Each element $v \in V$ exists in $K$ as a singleton set $\{v\}$. $K$ also contains 
$\sigma =\{v_0, \dots, v_k\} \subset V$ 
which is called a \emph{$k$-simplex} of dimension $k=\text{dim} \, \sigma = |\sigma| -1$.~e.g.~ edges ($k=1$), triangles ($k=2$), tetrahedra ($k=3$), etc. For every such $\sigma \in K$, all of $\sigma$'s subsets $\tau\subset\sigma$ are also included in $K$.  The dimension of the simplicial complex is the maximal dimension of its simplices:
$\text{dim} \, K=\max_{\sigma \in K} \text{dim} \,\sigma$. 
A graph $G=(V,E)$ is a 1-dimensional simplicial complex. Another well known special case is a triangle mesh $(V,E,F)$, which is a 2-dimensional embedded simplicial complex often with additional manifold conditions on the adjacencies of 2-simplices ({faces}) $F$.  A tetrahedral mesh is a 3-dimensional embedded simplicial complex. 

For a graph $G=(V,E)$, the incidence matrix  $A:E\rightarrow V$ where 
$A_{v,e}=1$ (or $-1$) depending on whether $e:v\rightarrow v'$ (or $e:v'\rightarrow v$ respectively) and zero otherwise, 
represents directed connectivity between vertices and edges. 
\emph{Boundary matrices} $\vartheta_k:\mathbb{R}[K_k]\rightarrow\mathbb{R}[K_{k-1}]$ extend this idea to higher dimensions and capture connectivity between (the vector space with real coefficients spanned by) the $k$-simplices $K_k$ and their bounding $K_{k-1}$ simplices. For convenience \emph{boundary operators} are  constructed by imposing an ordering on the vertices $V=K_0$, such that each $k$-simplex can be expressed by an \emph{ordered} list $\sigma=[v_0,\dots, v_k]$.

 The orientation of the simplices in a simplicial complex is dictated by the ordering imposed on the vertices, and the orientation of mesh elements given by the cyclic ordering of vertices. 
The boundary action is then applied on each simplex $\sigma$ according to
$
\vartheta_k(\sigma)=\sum_{i=0}^d(-1)^i\sigma_{-i},
$
where $\sigma_{-i}:=[v_0,\dots, \hat{v}_i, \dots, v_k]$ indicates the deletion of the $i$-th vertex from $\sigma$, resulting in a $(k-1)$-dimensional bounding simplex. 
The following figure illustrates a complex along with its three boundary matrices: white cells are zeros and grey cells are $\pm 1 $.

\vspace{.6em}
	\includegraphics[width=0.95\linewidth]{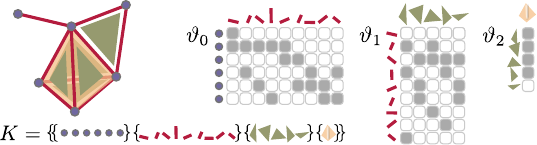}

\subsection{Hodge Laplacian as a generalization}\label{subsec:hodgeLapls}

\emph{Hodge Laplacians}~\cite{rosenberg1997laplacian} are differential operators that extend the notion of the well-known graph and mesh Laplacians to higher-dimensional simplicial complexes. It is defined, for each dimension $k$, as the sum of a two maps from $k$-simplices: one mapping down to $(k-1)$-simplices  and another mapping up to $(k+1)$-simplices. Each map is defined via its appropriate boundary matrix:
\begin{align*}
	L_k^{\text{rw}}\;\;&=& 
	L_k^{\text{down}}+L_k^{\text{up}} \;\; &=& 
	\vartheta_k^T W_{k-1}^{-1}\vartheta_k W_k + W_{k}^{-1}\vartheta_{k+1}W_{k+1}\vartheta_{k+1}^T.
\end{align*}
$W_k$ are diagonal matrices that contain a weight per $k-$simplex. The superscript clarifies that $L_k^{\text{rw}}$ is the \emph{random walk $k$-Hodge Laplacian} --  an antisymmetric linear map from $k$-simplices to $k$-simplices. It can be symmetrized, while preserving its spectral properties, as $L_k^w=W_k^{-1/2}\,L_k^{\text{rw}}\,W_k^{-1/2}$.  We direct the reader to key works~\cite{horak2013spectra, lim2020hodge} that present a thorough treatment of Hodge Laplacians and their spectra.
We refer to $L_k^{w}$ as the weighted $k$-Hodge Laplacian and to its unweighted (unit weights) version $L_k$ as simply the $k$-Hodge Laplacian.

Most variants of Laplacians used in graph- and mesh-processing may be derived as special cases of the weighted $k$-Hodge Laplacian. The graph Laplacian is a 0-Hodge Laplacian with unit weight on vertices $L^{\text{graph}}=L_0^{\text{rw}}=L_0^{\text{up}}=\vartheta_1W_1\vartheta_1^T$. The \emph{cotan Laplacian}, popular as the stiffness matrix in finite element methods, is the 0-Hodge Laplacian with different weights assigned to vertices and edges:
\begin{align*}
w_{v_i}=\sum_{\forall \sigma=\{v_i, v_j,v_k\}\in K_2}\mathcal{A}_\sigma/3 &,  \;\;\; &
w_{e_{ij}}=\frac{1}{2}(\cot \theta_{lij}+\cot \theta_{mij}).
\end{align*}
$\mathcal{A}_\sigma$ is the area of face $\sigma$, and $\theta_{lij}$ is the angle at vertex $v_l$ facing the edge $e_{i,j}=\{v_i,v_j\}$. Figure~\ref{fig:teaser} (left) depicts the impact of the choice of Laplacian on the coarsened \texttt{Fertility} triangle mesh. 

\paragraph{Hodge Spectra}

Every basis of the vector space $\mathcal{H}_k$ of a complex $K$ is spanned by a representative of equivalence classes of $k$-dimensional \emph{nontrivial loops}, and is isomorphic to the kernel of the $k$-Hodge Laplacian~\cite{eckmann1944harmonische}: $\ker L_k \simeq \mathcal{H}_k(K)$. The eigenvectors of $L_k$ corresponding to zero eigenvalues (the harmonic part of the spectrum) are minimal representatives of their respective homology classes.
An eigenvector corresponding to a non-zero eigenvalue of $L^\text{rw}_k$, must either be an eigenvector of $L_k^{up}$ or $L_k^\text{down}$ with the same eigenvalue. Furthermore, an element of the nullspace of $L_k^\text{rw}$ must be in the kernel of both of its components. Given an eigenpair $(\lambda_i,v_i)$ of $L_k^\text{down}$ then $(\lambda_i,\vartheta_k v_i)$ is an eigenpair of $L_{k-1}^\text{up}$~\cite{torres2020simplicial, horak2013spectra}.
 The \emph{Hodge decomposition} ties everything together, by expressing the space of $k$-simplices $K_k$ as a direct sum of \emph{gradients}, \emph{curls}, and \emph{harmonics}:
 \[K_k=\text{im}\vartheta_k^T\oplus\text{im}\vartheta_{k+1}\oplus\ker L_k. \]
We refer to several excellent introductions~\cite{lim2020hodge, chen2021helmholtzian} to Hodge decompositions.





\section{Method}
We propose a simple iterative algorithm for coarsening a simplicial complex based on two inputs: the fraction of simplices to be reduced and the portion of the input spectrum needing to be preserved. We use the latter to construct a quality function that is evaluated at each simplex. Then we greedily contract a simplex (or a group of simplices with low spectral-quality scores). We recalculate the quality function for simplices in the coarsened complex and iterate until the specified number of simplices have been reduced. 
\begin{figure}[htbp]
	\begin{centering}
	\includegraphics[width=.95\linewidth]{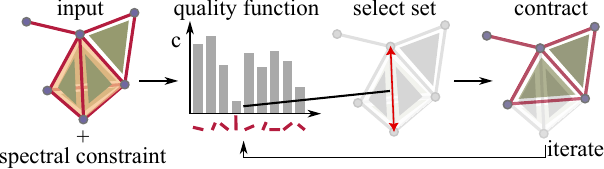}
	\caption{\label{fig:overview}An illustrated overview of our algorithm. }
			\end{centering}
\end{figure}

\subsection{Quality function} 
Each contraction (e.g.~edge collapse) can be defined as a projection $P_k:K_k\rightarrow \hat{K}_k$ mapping $k$-simplices in the input complex to those in the coarsened complex. Ideally, the eigenspace to be preserved should be perpendicular to that induced by the contraction. Intuitively the projection of the former onto the latter measures ``spectral leak'', with a large value indicating that the contraction leads to loss of fidelity with respect to the specified target spectrum. 

Let the eigenspace to be preserved be represented by eigenvectors $U$ and eigenvalues $S$ (diagonal matrix) of the Hodge Laplacian $L_k^\text{w}$ (or $L_k$) of the input complex, and let $A_k = U\sqrt{S^+}$ be the preserved subspace. Since $P$ and its pseudoinverse $P^+_k = P_k^T D^{-2}$, where $D$ is a diagonal (normalizing) matrix containing row norms, map in opposite directions to and from the output complex, the operator $\Pi = P_k^+P_k$ projects a signal defined on the fine complex down to the coarsened complex and back up to the fine complex.

Ideally, we seek contractions where $\Pi_k A_k$ is identical to $A_k$ by minimizing $\norm{\Pi_k A_k-A_k}_{\bullet}$. Our quality function $c_k: K_k \rightarrow \mathbb{R}$ 
\begin{equation}
	\label{eq:qf}
	c_k = \| \Pi_k^\perp \; A_k^{\ell} \|_L, 
\end{equation}
where the perpendicular projection $\Pi_k^\perp = (\mathbb{I} - \Pi_k)$, quantifies spectral error. Here $\norm{x}_L=\sqrt{x^TLx}$ denotes the $L$-norm. The above definition of $c$ in terms of $\Pi^\perp_k$ allows for localized error computations, facilitating parallelization, since the nonzero entries of $\Pi^\perp_k$ only refer to simplices affected by the contraction. This quality function, has been analyzed to have desirable theoretical properties when applied to graph simplification and general positive semidefinite matrices~\cite{loukas2019graph}. 

\subsection{Contraction}
The first step to performing contractions, or setting equivalences of multiple simplices, is to identify a set $\Phi$ of candidate contraction sets $\phi_i\in\Phi$. We execute edge collapses by setting each $\phi_i$ to be an edge. More complicated contractions, such as collapsing stars around vertices, may be performed by  identifying the relevant $\Phi = \{\phi_i\}$.

Then, we evaluate the quality function over each candidate contraction set and greedily contract the set $\phi^*$ with the minimum quality. This involves identifying all the simplices in $\phi^*$ to one, target, vertex.  As a result, the complex is coarsened while best preserving the spectral band $(U,S)$ of $L_k$.  In practice, we apply such  contractions iteratively until a specified ratio of simplices has been contracted, with some additional bookkeeping at each level $\ell$: starting from  $A^{\ell=0}_k = U\sqrt{S^+}$, we update the target subspace and the corresponding projection matrices $P^{\ell}_k$ and $\Pi^{\ell}_k$, with $A_k^\ell$ tracking the evolution of the desired spectrum at each iteration.
 
\subsection{Multiple Laplacian subspaces}
The quality function (Equation~\ref{eq:qf}) can be adapted so that the target subspace is shaped via multiple spectral bands. We define cost functions $c_k^{\ell,\,\nu}$ at level $\ell$ independently for $B$ different spectral bands $\nu=1,2,\cdots, B$.  Each of these is associated with a potentially different Laplacian. The aggregate cost function can then be tailored  based on the specific downstream application as  
 $q_{\text{agg}}: \mathbb{R}^B \rightarrow \mathbb{R}$ 
 to obtain the final quality function
 $c^{\ell}_k=q_{\text{agg}}(\{c^{\ell,\,\nu}_k\}), \;\; \nu\in[0,1,\cdots,B].$
 
In our experiments in Section~\ref{sec:eval}, where $L_0$, $L_1$ and $L_2$ are considered in tandem, we simply average the contributions of the different Laplacians $q_\text{agg}({c^v_k})=\frac{1}{B}\sum_{k,v} c^v_k$.
Algorithm~\ref{algo:quality} shows our generalized coarsening procedure by assembling the above stages. 

\let\oldnl\nl
\newcommand{\nonl}{\renewcommand{\nl}{\let\nl\oldnl}}
\begin{algorithm}[h]
	 \SetKwInOut{Input}{inputs}
	\SetKwInOut{Output}{output}
	
	\Input
	{		
		\par 
		\Indp
		$K = \{K_k\} $\tcp*[r]{high-resolution simplicial complex}
		\par
		\nonl $\rho$ \tcp*[r] {fraction of simplices to reduce} 
		\par
		\nonl $\{(S_0,U_0),\dots,(S_B,U_B)\}$ \tcp*[r]{target subspaces to preserve} 
		\par 
		\nonl $L= \{L_0, \dots, L_B\}$ \tcp*[r]{Laplacians corresponding to subspaces}
	} 

	\Output
	{
		\par 
		\Indp 
		$\hat{K}$ \tcp*[r]{Coarsened complex}
		\nonl $P_k$ \tcp*[r]{projection matrices (fine to coarse)}
	}
	\vspace{.25em}
	{\hrule}
	\vspace{.45em}
	$\hat{K} \gets K$\;
	$\ell \gets 0$ \;
	
	\While{ $1 - |\hat{K}|/|K| < \rho$ }
	{
		Define $\Phi$ \tcp*[r]{say, set of all edges of $\hat{K}$}
			
		\ForEach{ subspace $(S_\nu, B_n)$}
		{ 
			\eIf{$\ell=0$}{ 
				$M^\ell_\nu\gets U_\nu S_\nu^{+1/2}$\;
				$A^\ell_\nu\gets M^\ell_\nu$\;
			}{
				$L^\ell_{\nu} \gets P^{\ell \mp} L^{\ell-1}_{\nu} P^{\ell +}$				\tcp*[r]{coarsened Laplacian}
				$M^\ell_\nu\gets P^\ell M^{\ell-1}_\nu$\;
				$A^\ell_\nu\gets M^\ell({M^\ell_\nu}^T L^\ell_\nu M^\ell_\nu)^{+1/2}$ 
				\tcp*[r]{coarse target subspace }
			}
		}
	
		\ForEach{set of simplices $\phi\in\Phi$}{
			$\Pi^\perp\gets (\mathbb{I}-P^+P)$\;
			\ForEach{ subspace $(S_\nu, B_{\nu})$ }{ 
				$c^{\ell,\,\nu}_\phi\gets \norm{\Pi^\perp A^\ell_\nu}$ \;
			}
			$c^\ell_\phi\gets q_\text{agg}(\{c^{\ell,\,\nu}_\phi\})$ \tcp*[r]{user-specified aggreg. function}
		}

		$(\hat{K}, P^{\ell}) \gets \mathrm{Contract}(\hat{K}, \argmin_\phi\{c^\ell_\phi\})$\;
		$\ell \gets \ell + 1$\;
	}
		
	\caption{Iterative generalized spectral coarsening}
	\label{algo:quality}
\end{algorithm}

\subsection{Implementation details}
\paragraph{Terms of the Hodge Laplacian} Due to the interplay of spectra of $L_k$ and $L_{k+1}$ discussed in Section~\ref{subsec:hodgeLapls}, constraints on Laplacians for multiple $k$ can potentially introduce `spectral conflicts'. We avoid this by considering only the $L_k^\text{up}$ components of the chosen Laplacians for each $k$. For wide spectral bands and appropriate choice of $k$, the spectral region of interest will be a subspace of the spectrum of $L_k^\text{up}$ and thus preserved.

\paragraph{Building coarsening matrices $P_k$} During contraction, let $z$ be the index of a contracted simplex in $K$, $t$ be the index of the target simplex in $\hat{K}$ and   $|\phi|$ bet the cardinality of the candidate family being contracted. $P_k$ is then a simplicial map $K_k\rightarrow\hat{K}_k$ where every $k-$simplex of $K$ maps to a valid simplex of $\hat{K}$, or to zero. For consistency, it is required that $\hat{\vartheta}_kP_k = P_{k-1}\vartheta_k$. To satisfy desirable spectral approximation guarantees~\cite{loukas2019graph} and for $P^+=P^TD^{-2}$ to hold, we set the elements of each contraction set as $P_{t,z}=\frac{1}{|\phi|}$, for all $z\in \phi$, where $\phi\mapsto t$.  We resolve the ambiguity of the choice of target simplex by consistently selecting the target simplex based on its index.~e.g.~for edge contractions, both vertices are merged into one with the smaller index. 


\paragraph{Choosing candidate families $\Phi$} Our algorithm is agnostic of the choice of combinations of simplices $\Phi$ to be contracted. We tested with various candidate families: edges (pairs of vertices), faces (triplets of vertices) and more general vertex neighborhoods consisting of closed-stars. Larger contraction sets result in aggressive coarsening at each iteration which leads to larger spectral error. All results in this paper use only edge collapses. This also simplifies comparison to related work (which are restricted to edge collapses).

\paragraph{Harmonic subspaces} For $k>0$, the harmonic portion of the spectrum of $L_k$ is non-trivial and encodes information about non-trivial cycles called \emph{homology generators}. A homology group $\mathcal{H}_k$ is a vector space of cycles that are not bounding higher dimensional simplices and therefore manifest as the null space of $L_k$.  It is often desirable to preserve these eigenvectors, despite their coresponding zero eigenvalue, to maintain topological consistency. In practice, for all experiments we use modified eigenvalues $\tilde{S}=\mathbb{I}+S$ when dealing with spectral of higher-dimensional Laplacians. 


\paragraph{Local Delaunay vertex position optimization} Edge contractions may produce self-intersections, particularly if a fixed vertex-placement policy is followed, say, at the midpoint of an edge. While there are methods to avoid this effectively~\cite{sacht2013consistent}, we adopted a simple vertex positioning scheme for tetrahedral meshes to prevent intersections. 
\begin{center}
	\includegraphics[width=.95\linewidth]{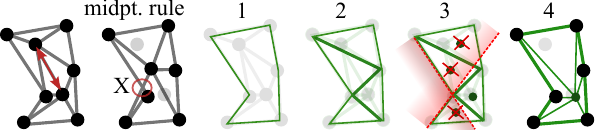}
\end{center}
Our scheme operates in four steps: (1) identify the faces in the \emph{link} of the edge being collapsed; (2) construct the Delaunay tetrahedralization of the vertices of these faces; (3) choose barycenters of the new tetrahedra that are on the same side of the faces in 1 as the corresponding vertex of the edge being collapsed; (4) the output vertex is the centroid of all vertices that pass the test in 3. If no vertices pass the test in 3 then we do not perform the collapse.

\newcommand{\rootwidthTET}{0.249}
\begin{figure}
	\begin{center}
		\begin{tabular}{@{}c@{}c@{}c@{}c@{}}
			Ref . (11385 v.) & 5000 verts. & 3000 verts. & 1000 verts. \\
			\includegraphics[width=\rootwidthTET\columnwidth]{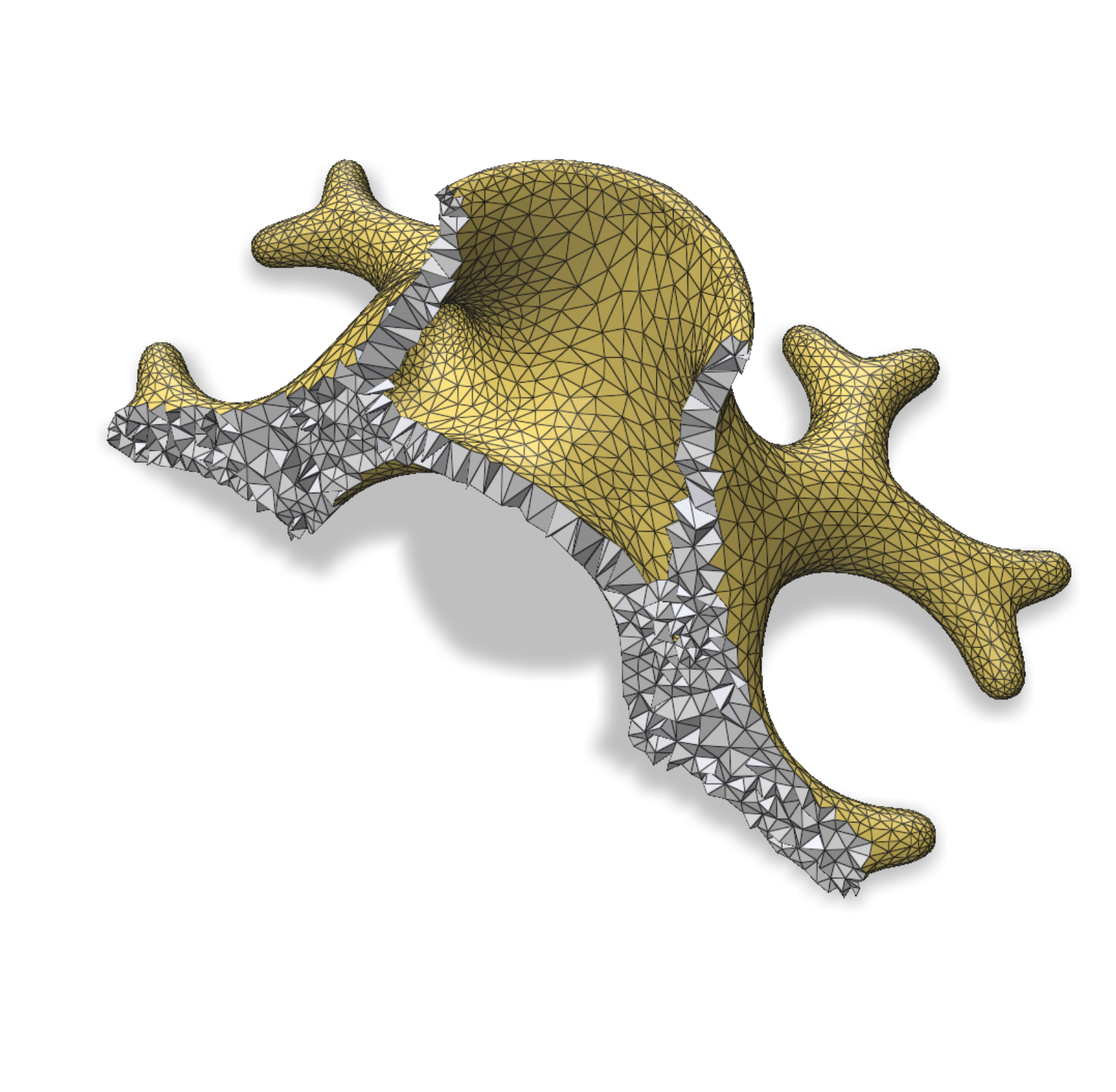} &		
			\includegraphics[width=\rootwidthTET\columnwidth]{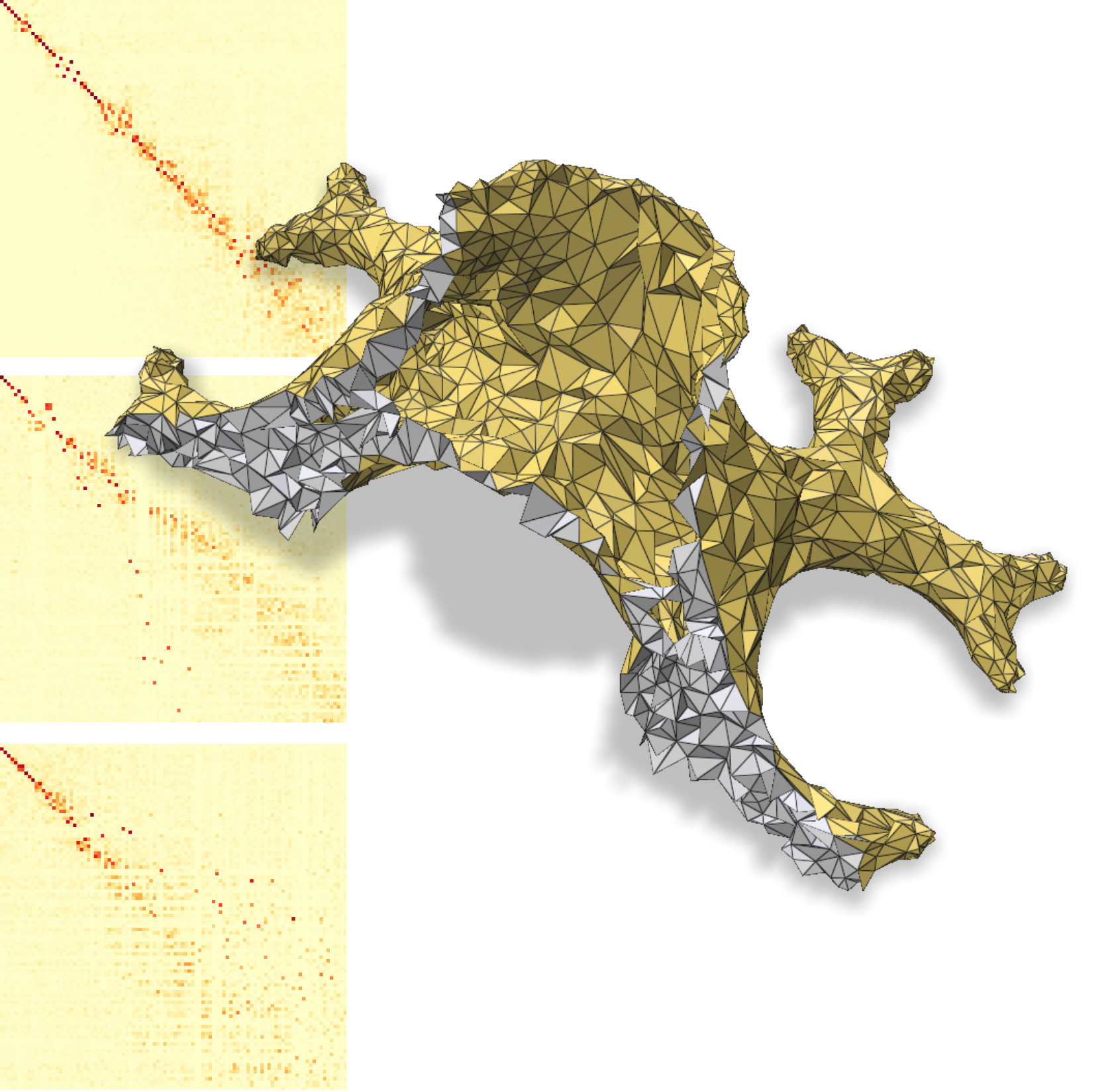} &
			\includegraphics[width=\rootwidthTET\columnwidth]{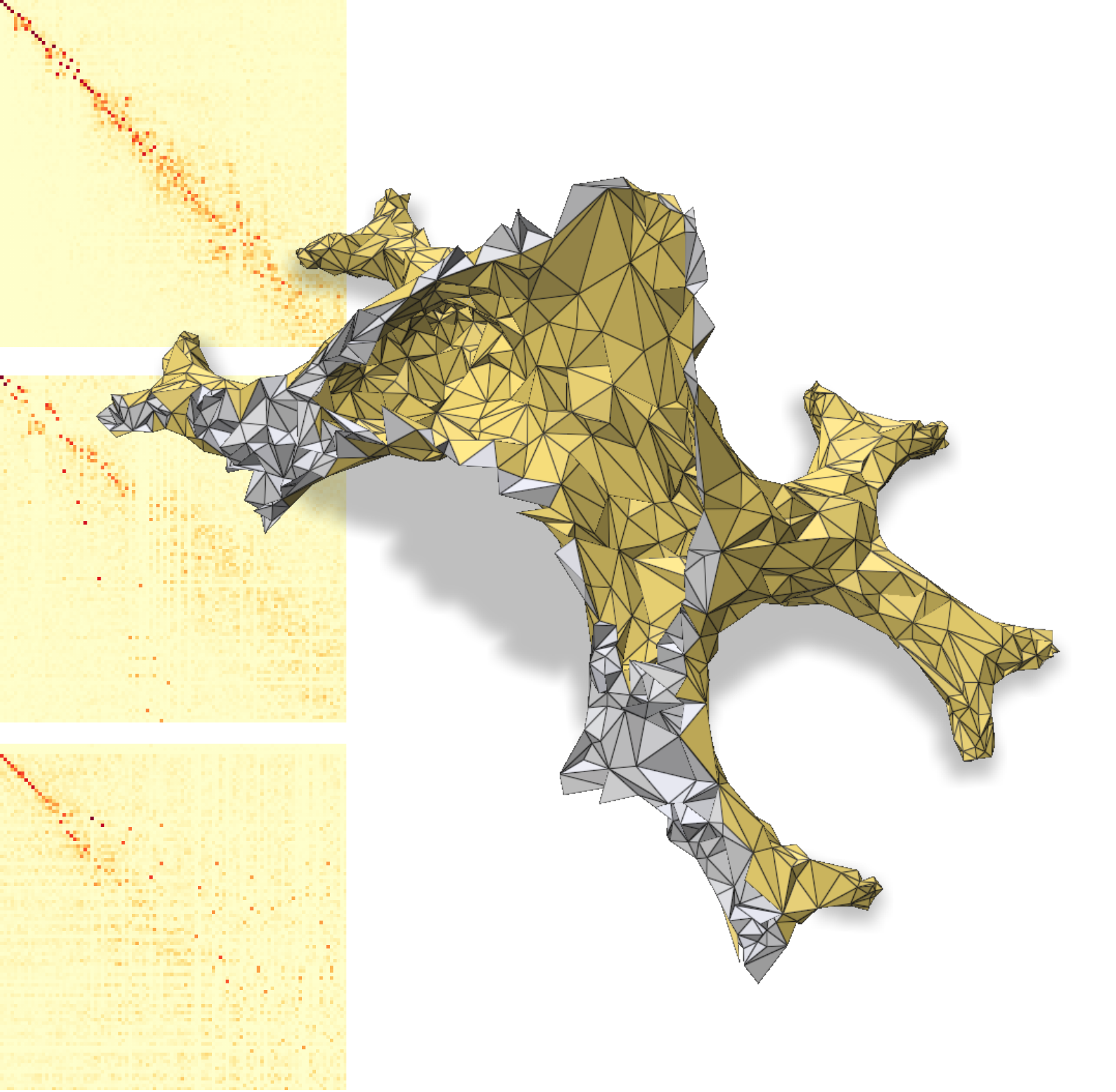} &
			\includegraphics[width=\rootwidthTET\columnwidth]{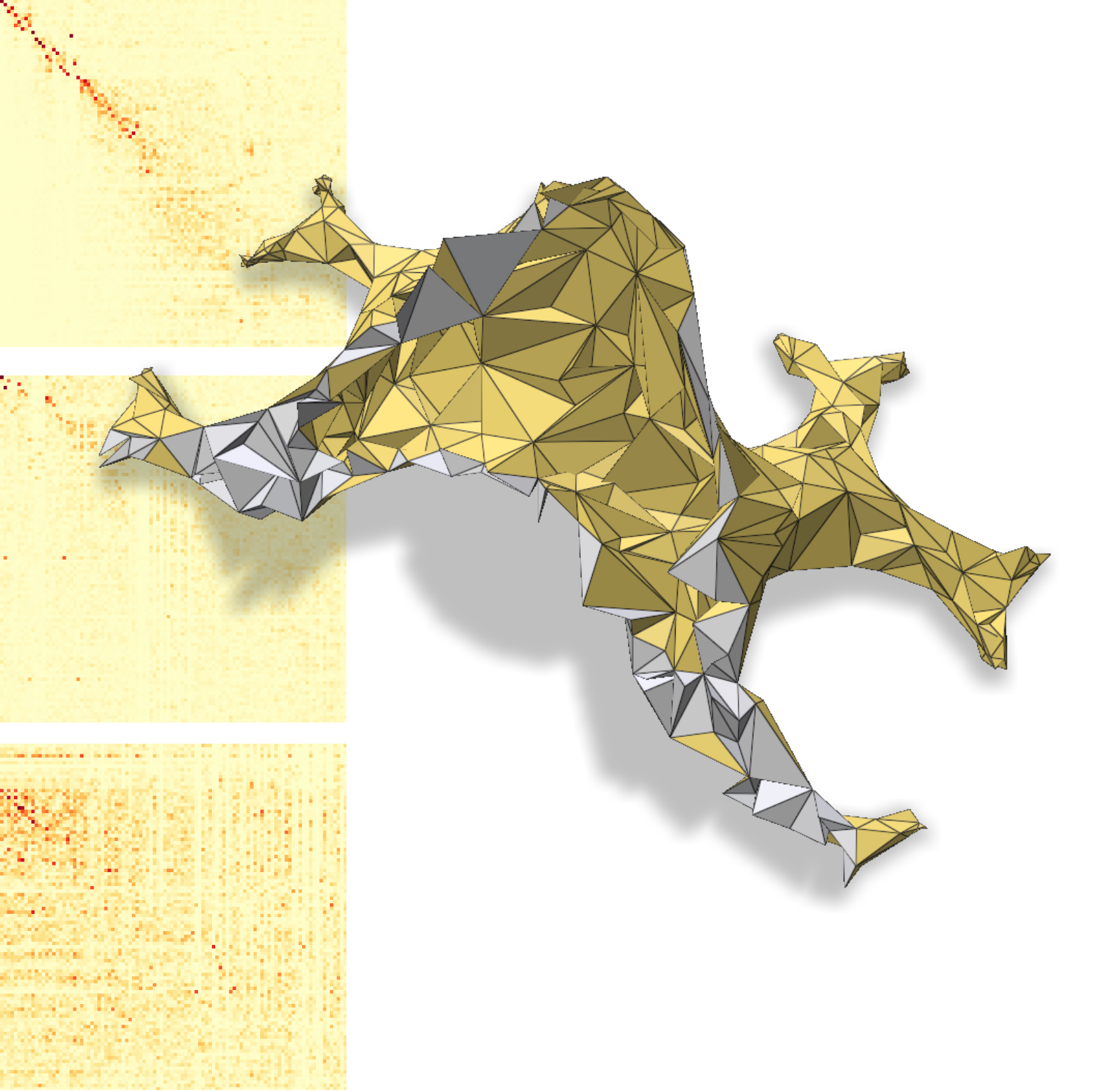}
		\end{tabular} \\
			\begin{tabular}{@{}@{\;\;\;}c@{\;\;\;}c@{\;\;\;}  >{\columncolor[gray]{0.96}}cc>{\columncolor[gray]{0.96}}cc>{\columncolor[gray]{0.96}}cc@{}}
				\toprule
				& $k$ & \small{$\|\cdot\|_{\Pi^{\perp}}$} & \small{$\|\cdot\|_{L_\text{c}}$} & \small{$\|\cdot\|_{C_\text{conf}}$ } & \small{$\|\cdot\|_\text{sub}$} &\small{ $\|\cdot\|_{\Theta}$ } & \small{$\|\cdot\|_{\lambda}$} \\
				\midrule
				\multirow{3}{*}{\rotatebox{90}{5000 v.} }
				& 0 &                        39.126 &                      1.023 &                       7.855 &                   0.010 &               13.8 &                 9.07 \\
				{}   &1 &                        73.351 &                      0.007 &                       1.251 &                   0.183 &                8.4 &                 5.165 \\
				{}   &2 &                        83.015 &                      0.003 &                       1.241 &                   0.174 &                2.95 &                 2.52 \\
				\midrule
				\multirow{3}{*}{\rotatebox{90}{3000 v.}} & 0 &                        61.272 &                      3.298 &                      13.506 &                   0.002 &                8.17 &                19.51 \\
				{} & 1 &                        91.813 &                      0.013 &                       1.723 &                   0.381 &                2.48 &                10.138 \\
				{} & 2 &                        97.56 &                      0.004 &                       1.583 &                   0.489 &                0.44 &                 4.89 \\
				\midrule
				\multirow{3}{*}{\rotatebox{90}{1000 v.}} & 0 &                        87.90 &                      8.266 &                      26.511 &                   0.009 &                2.45 &                55.53 \\
				{}   & 1 &                        97.72 &                      0.026 &                       2.804 &                   0.532 &                0.75 &                23.139 \\
				{}   & 2 &                        99.60 &                      0.011 &                       2.434 &                   0.709 &                0.07 &                11.34\\
				\bottomrule
			\end{tabular}
			\caption{\label{fig:specerrorincrease} Spectral error increases controllably as the reference tetrahedral mesh (top left) with $11,385$ vertices is progressively coarsened while preserving  the spaces spanned by the first $100$ eigenvectors of three Laplacians: $L_k, \, k=0,1\, \mathrm{and} \, 2$.}
		\end{center}
	\end{figure}
	
	\begin{figure}[htbp]
		\begin{center}		
			\begin{tabular}{@{}l@{\;\;\;}c@{\;\;\;} >{\columncolor[gray]{0.96}}c@{\;\;\;}c@{\;\;\;} >{\columncolor[gray]{0.96}}c@{\;\;\;}c@{\;\;\;} >{\columncolor[gray]{0.96}}c@{\;\;\;}c@{}}
					\toprule
					&  $k$ & \small{$\|\cdot\|_{\Pi^{\perp}}$} & \small{$\|\cdot\|_{L_\text{c}}$} & \small{$\|\cdot\|_{C_\text{conf}}$ } & \small{$\|\cdot\|_\text{sub}$} &\small{ $\|\cdot\|_{\Theta}$ } & \small{$\|\cdot\|_{\lambda}$} \\
					\midrule
					\multicolumn{8}{c}{ triangle mesh } \\
					ours & 0           &                         9.94 &       $\mathbf{5.6\,e5}$  &               $\mathbf{2393}$ &    $1.5\, e{4}$ &               37.314 &               $\mathbf{198}$ \\
					baseline & 0&                         $\mathbf{6.82}$ &       $2.6 \, e7$ &                    4738  &               $\mathbf{2311}$ &               $\mathbf{21.127}$ &              5648 \\		
					ours & 1 &             31.35 &                      $\mathbf{0.008}$ &                       $\mathbf{0.787}$ &                   0.350 &                $\mathbf{0.805}$ &                 \textbf{0.590} \\
					baseline & 1 &                        $\mathbf{27.82}$ &                      0.018 &                       0.827 &                   $\mathbf{0.159}$ &                2.555 &                 0.635  \Bstrut\\
					\multicolumn{8}{c}{tetrahedral mesh} 
					\\
					ours & 0 &                        22.379 &                    142 &                     130 &    $3.1 \, e{4}$ &               24.58 &                 1.827 \\
					ours& 1&                        83.90 &                      0.030 &                       2.206 &                   0.174 &                4.999 &                 9.581 \\
					ours & 2&                        91.14 &                      0.007 &                       1.696 &                   0.287 &                1.289 &                 1.938 \\
					\bottomrule
				\end{tabular}
				\caption{\label{tab:spectralerror} A comparison of spectral error resulting from our method to the baseline on a Suzanne triangle mesh (top portion) using different metrics. The bottom portion reports spectral errors on a Fertility tetrahedral mesh.  }
			\end{center}
		\end{figure}

\section{Results} \label{sec:eval}
Unless otherwise specified, we use a common low-pass constraint as the default for experiments: to preserve the space spanned by the first $100$ eigenvectors of the specified Laplacian(s). 

\subsection{Meshes}
\paragraph{2D Baseline}
To enable comparisons of $L_k, \; k=1$ for triangle meshes we extend previous work~\cite{lescoat2020spectral} as a baseline. Their method minimizes spectral error
$ E=\norm{PM^{-1}LF-\tilde{M}^{-1}\tilde{L}PF}_{\tilde{M}}^2,$ where $P$ is a coarsening projection matrix, $M$ is a mass matrix, $L$ is a differential operator, and $F$ is the spectrum of interest as a matrix of eigenvectors. A tilde above the respective notations denotes their coarsened versions. From their output coarsening matrix $P_0$ we additionally infer coarsening matrices $P_1$ and $P_2$ which operate on the space of edges and faces respectively. However, they \emph{do not consider higher dimensional mappings in their error metric}. Since they only consider one Laplacian $L_0^{\mathrm{cot}}$, we include this as one of the targets in all our comparisons, unless stated otherwise. We visualize spectral preservation via functional maps~\cite{liu2019spcoarsegeomops,lescoat2020spectral, chen2020chordal}
$ C=U_c^T P U$ which is a diagonal matrix when the input and output spectra match perfectly. 
 Unfortunately, it is not straightforward to compare with prior non-spectal tetrahedral coarsening methods, or to extend previous spectral coarsening methods to operate on tetrahedral meshes (Section~\ref{sec:relatedwork}).

\paragraph{Quantitative comparison}
We visualize coarsened triangular (Figure~\ref{fig:eigenvectors2D}) and tetrahedral (Figure~\ref{fig:eigenvectors3D}) meshes along with a few eigenvectors (as heat maps on vertices).  The results are reassuring that for the special case of coarsening triangle meshes our algorithm produces similar spectral results to previous work. Using $L_1$ subtly improves the preservation of structures such as the jaw-line, the mouth and the nose of the \texttt{Suzanne} model (Figure~\ref{fig:eigenvectors2D}). The boundary loops around the eyes (eigenvectors of the null space of $L_1$) aim to preserve their original size compared to when only $L_0$ is used. 
%
The bottom row of Figure~\ref{fig:eigenvectors2D} contains quantitative comparisons of our eigenvalues against the baseline and reference. It appears that the approximation is good for $L_0$, but curiously spectral divergence is observed for our method and the baseline in $L_1$ . 

This is also observed on a tetrahedral mesh (\texttt{Fertility} model) as shown in the bottom row of Figure~\ref{fig:funcmaps}. However,  the profile of eigenvalues for $L_1$ and the eigenvalues of $L_2$ are approximated well. We also show functional maps to support these observations. The Laplacians considered were $L_0^{\mathrm{cot}}$ and $L_1$ for the triangle mesh (2D), and additionally $L_2$ for the tetrahedral mesh. 
We report errors in Table~\ref{tab:spectralerror} using the following error metrics:  
a local volume-preserving measure $\|C\|_{\Pi^{\perp}}=\norm{C^TC-\mathbb{I}}_F^2$, 
an isometry measure $\|C\|_{L_\text{c}}=\norm{S_cC-CS}_F^2/\norm{C}_F^2$, a conformality measure $\|C\|_\text{conf}=\norm{C^TSC-Sc}_F^2$, a subspace approximation measure $\|C\|_\text{sub}=|\norm{S^{1/2}U^T \Pi U S^{-1/2}}_2-1|$, 
subspace alignment $\|C\|_\theta=\norm{\sin\Theta(U, P^+U_c)}_F^2$ and relative eigenvalue error $\|S\|_{\lambda}=\norm{(S-Sc)/S}_2$. 

The above results indicate comparable performance against the baseline for for triangle meshes and the added capability of handling higher dimensional Laplacians for tetrahedral (3D) meshes. Additional examples are included in the supplementary material.  We illustrate the evolution of spectral errors for the \texttt{Tree root} model over three levels of coarsening (Figure~\ref{fig:specerrorincrease}).

\subsection{Simplicial Complexes}\label{sec:complexes}
We compare our coarsening algorithm to the state-of-the-art \emph{edgecollapser} of the Gudhi library~\cite{gudhi}, that guarantees homology preservation, and a simple baseline where edges are collapsed at random.

\paragraph{Band-pass filtering}
We demonstrate the versatility of our method by coarsening an input complex with two different spectral bands to be preserved: $\beta_1+1$ lowest, and 10 largest eigenpairs of $L_1$. Here, $\beta_1$ is the betti number (rank of the homology group $\mathcal{H}_1$).  Figure~\ref{fig:bandpass} presents these results along with the functional maps $C$ as explained above. As expected (by design), Gudhi exactly preserves the homology rank $\beta_1$, with no regard for other features. On the contrary, our method can be tuned to a spectral band of choice. Either the nullspace-encoded harmonic information (first row of functional maps) or the high frequency band(second row of maps).

\paragraph{Homology preservation}
We constructed a diverse dataset of 100 simplicial complexes~\cite{keros2022dist2cycle} with non-trivial topology by randomly sampling 400 points repeatedly on multi-holed tori, and subsequently constructing \emph{alpha} complexes at various thresholds.  It contains from zero up to 66 homology cycles. On this dataset, we computed mean spectral error metrics (Table~\ref{tab:homologyMetrics}), alongside a homology preservation error $E_\beta = |\beta_1^{\mathrm{fine}}-\beta_1^{\mathrm{coarse}}|$. Intuitively, this is the error in the number of 1-cycles destroyed by coarsening. The number of simplices reduced by Gudhi was consistent across both experiments.  We ran two variants of our method which preserved the first $30$ eigenpairs of $L_1$, reducing the input complexes by a factor of 0.8, and the first $\beta_1+1$ eigenpairs of $L_1$, matching the target number of vertices to the result of Gudhi. The results are summarized in Table~\ref{tab:homology} with standard deviations in parentheses. Gudhi, as designed, preserves $\beta_1$ exactly but exhibits spectral leak elsewhere. With random contractions, the leak is amortized across spectral bands but it destroys about 1 cycle on average. Our method is controllable, highlighting that it can be particularly effective if the spectral constraints are known for a particular application.

\newcommand{\WDev}[2]{\makecell{#1\\(#2)}}

\begingroup
\renewcommand{\arraystretch}{2.1} 
\begin{table}
	\begin{tabular}{@{}c|@{\;\;\;}c@{\;\;\;} >{\columncolor[gray]{0.96}}c@{\;\;\;}c@{\;\;\;} >{\columncolor[gray]{0.96}}c@{\;\;\;}c@{\;\;\;} >{\columncolor[gray]{0.96}}c@{\;\;\;}c@{}}
			\toprule
			{k} & {} & {$\rho$} & \small{$\|\cdot\|_{L_\text{c}}$} & \small{$\|\cdot\|_{\Pi^{\perp}}$} & \small{$\|\cdot\|_\text{sub}$} & \small{$\|\cdot\|_{\lambda}$} &           $E_{\beta}$ \\ 
			\hline
			30 & Gudhi  & 0.9 & \WDev{3.84} {4.2} & \WDev{29.1} { 2.9}& \WDev{2.5}{16.8}& \WDev{71.5}{664}& \WDev{0.0}{0.0}  \\	
			 & Ours & 0.8 & \WDev{\textbf{0.49}} {0.5} & \WDev{\textbf{8.98}} {3.5}& \WDev{1.52}{12.2}& \WDev{\textbf{2.76}}{10.5}& \WDev{0.07}{0.2}  \\			
			 & Random& 0.8 & \WDev{3.08} {2.2} & \WDev{20.8} {3.5}& \WDev{0.32}{0.4}& \WDev{400000}{3 e6}& \WDev{.98}{1.2}  \\				
			\midrule
			$\beta_1+1$&  Gudhi & 0.9 & -  & \WDev{2.94} {1.0}& \WDev{0.91}{0.2}& - & \WDev{0.0}{0.0}  \\			
			&  Ours  & 0.9 & -  & \WDev{\textbf{1.78}} {1.2}& \WDev{\textbf{0.78}}{0.3}& - & \WDev{0.21}{0.6}  \\			
			& Random  & 0.9 & -  & \WDev{{2.76}} {1.1}& \WDev{{0.88}}{0.2}& - & \WDev{1.29}{1.4}  \\			
			\bottomrule
		\end{tabular}
	\caption{\label{tab:homologyMetrics}Spectral approximation metrics averaged over 100 complexes. Two sets of executions of our algorithm preserve the first $30$ and the first $\beta_1+1$ dimensions of the eigenspace respectively.  The various error metrics (columns) are described in Section~\ref{sec:complexes}. Standard deviations are shown in parentheses. }
	\label{tab:homology} 
\end{table}
\endgroup
	
\newcommand{\cmplxwidth}{0.2}
\newcommand{\matwidth}{0.2}
\begin{figure}
	\begin{center}
		\begin{tabular}{@{}r@{}c@{}c@{}c@{}c@{}c@{}}			
			\multicolumn{1}{c}{Input} & Random & Gudhi & Ours (low) & 
			Ours (high) \\
			\includegraphics[width=\cmplxwidth\linewidth]{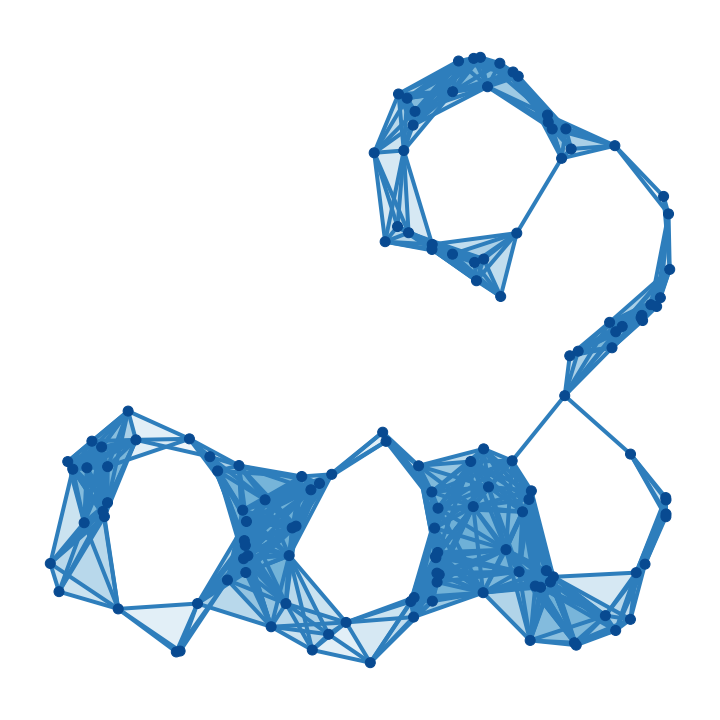} &
			\includegraphics[width=\cmplxwidth\linewidth]{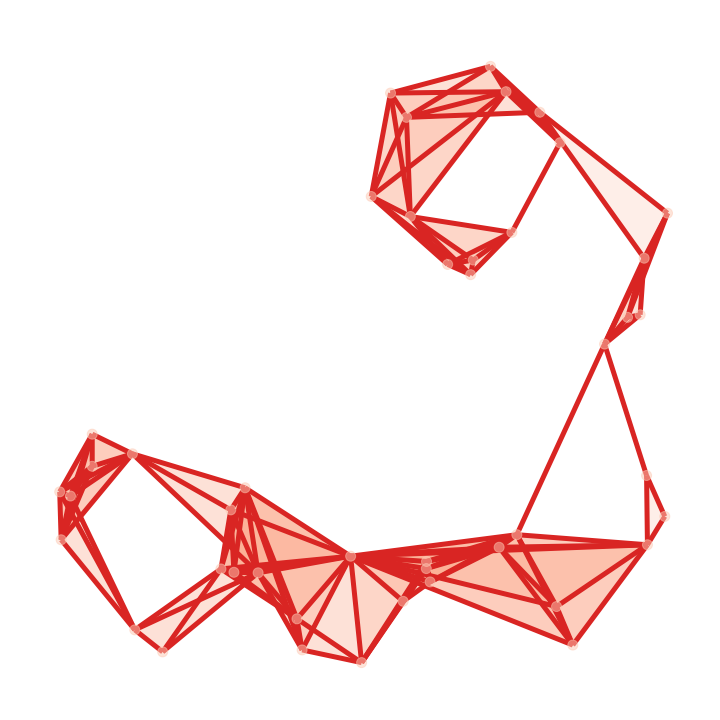} &
			\includegraphics[width=\cmplxwidth\linewidth]{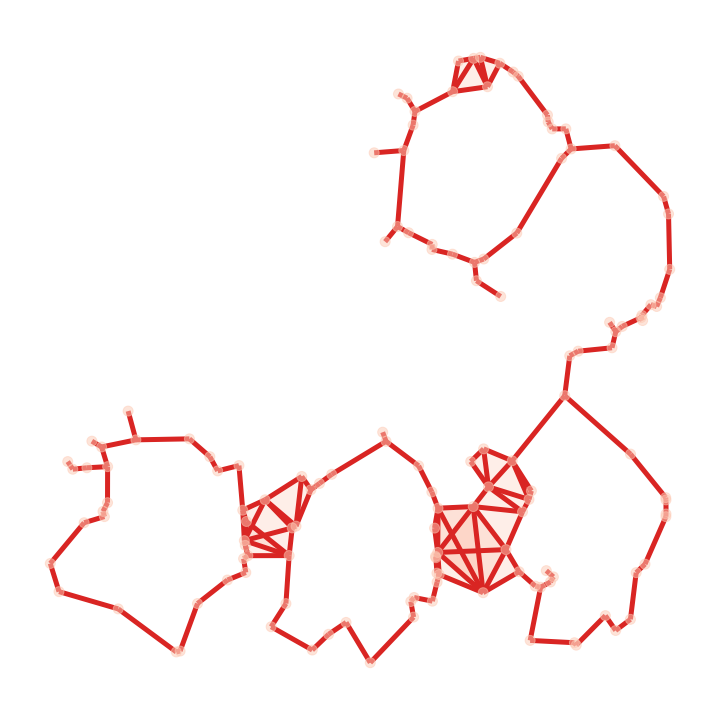} 
			&
			\includegraphics[width=\cmplxwidth\linewidth]{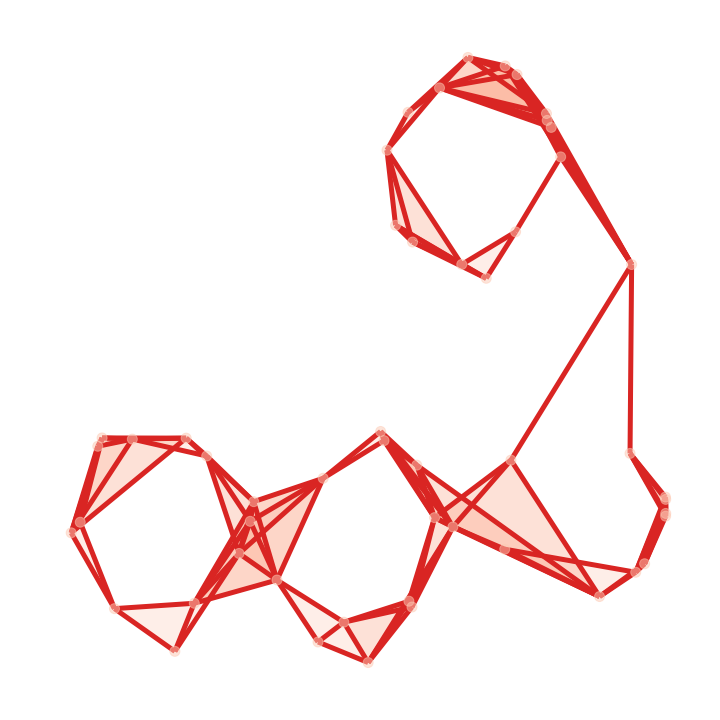} 
			& 
			\includegraphics[width=\cmplxwidth\linewidth]{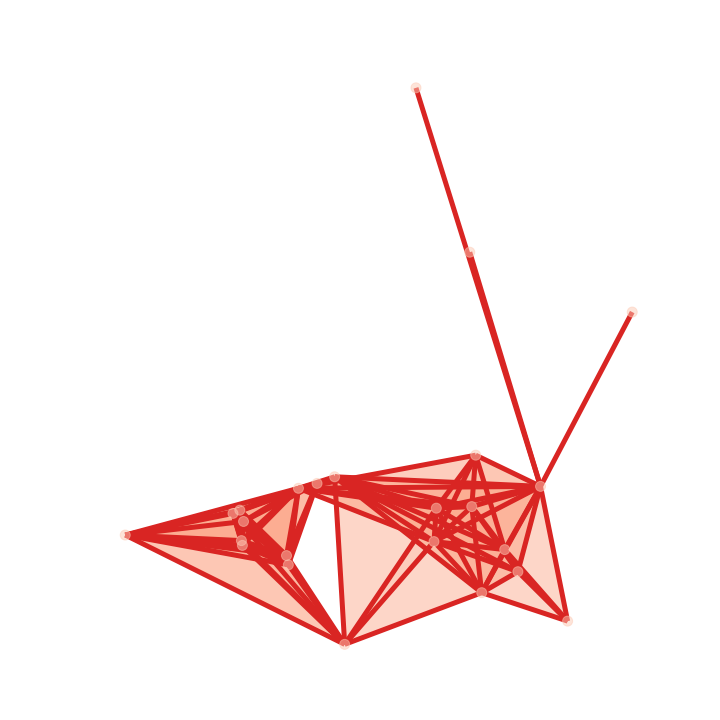} \\
			\rotatebox{90} {0-10} &\includegraphics[width=\matwidth\linewidth]{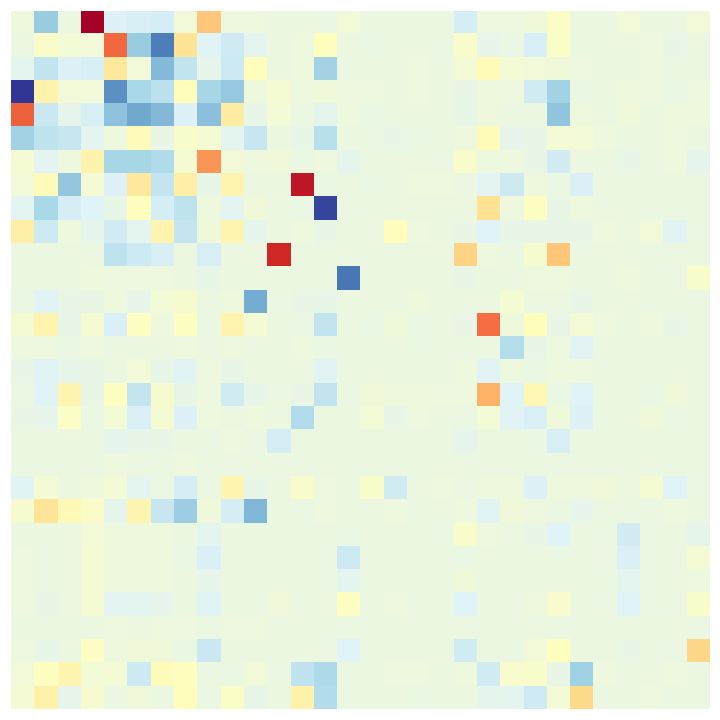} & 
			\includegraphics[width=\matwidth\linewidth]{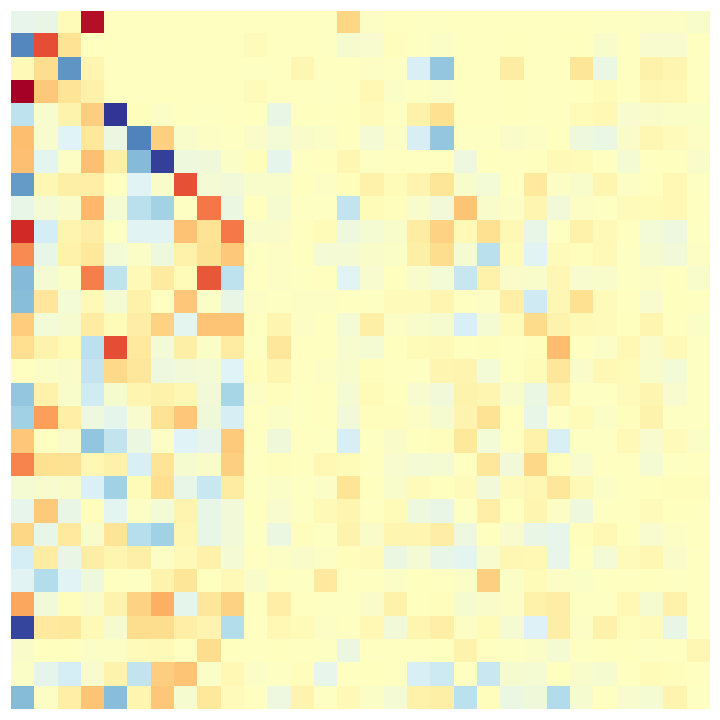}&
			\includegraphics[width=\matwidth\linewidth]{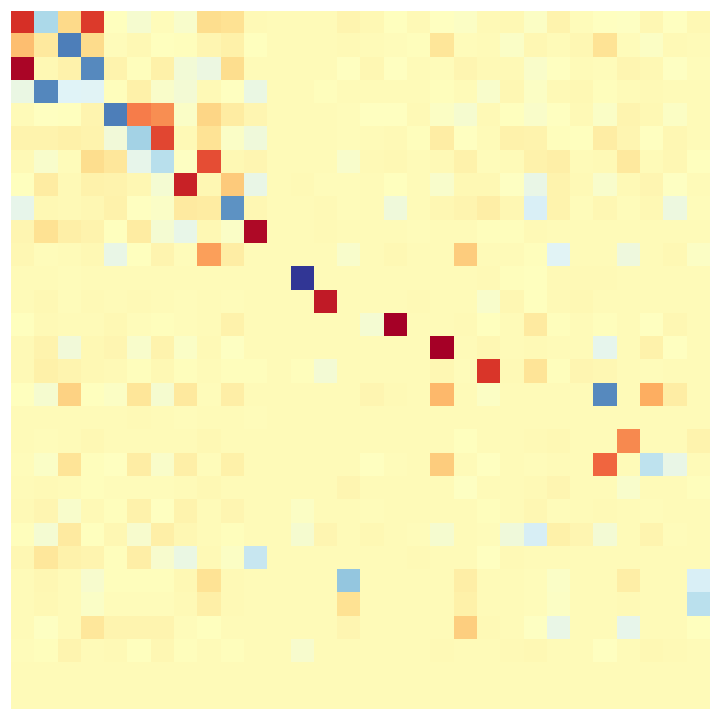}&
			\includegraphics[width=\matwidth\linewidth]{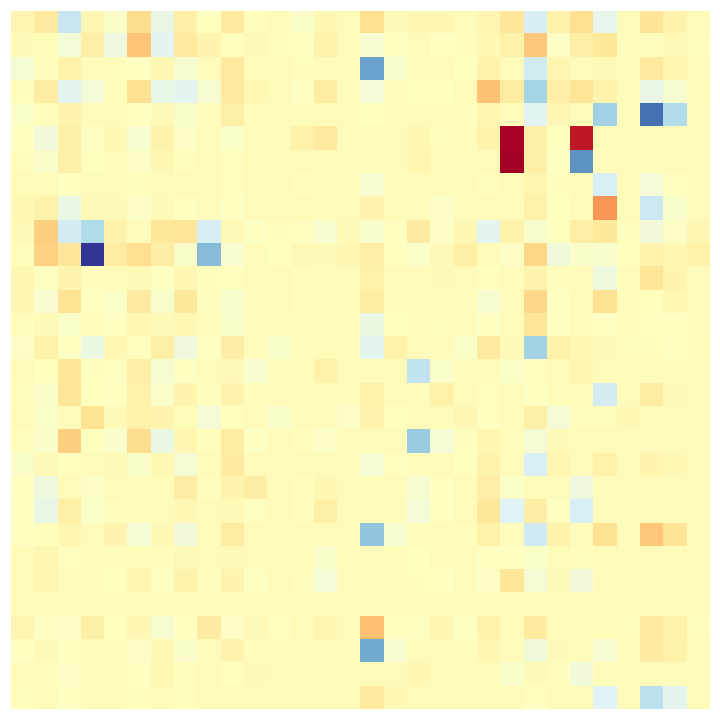} \\
			\rotatebox{90} {790-800} &\includegraphics[width=\matwidth\linewidth]{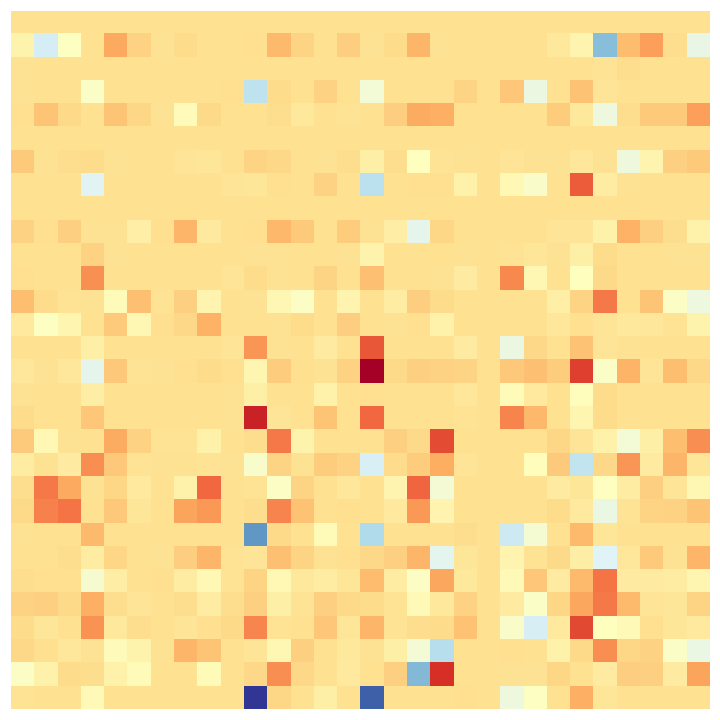} & 
			\includegraphics[width=\matwidth\linewidth]{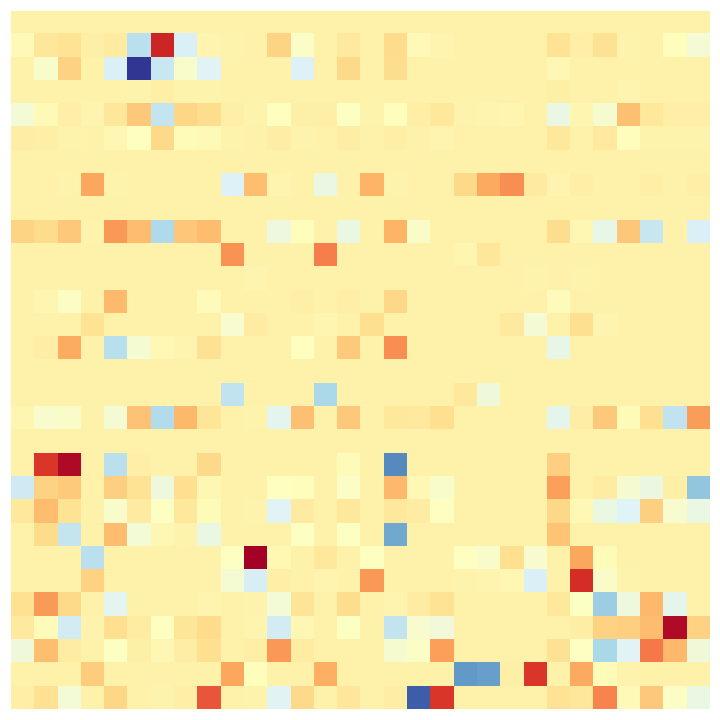}&
			\includegraphics[width=\matwidth\linewidth]{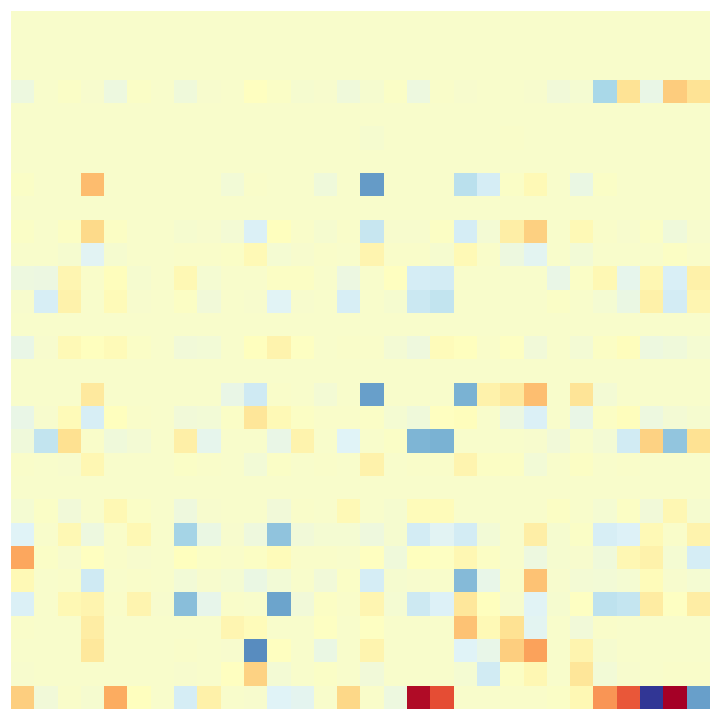}&
			\includegraphics[width=\matwidth\linewidth]{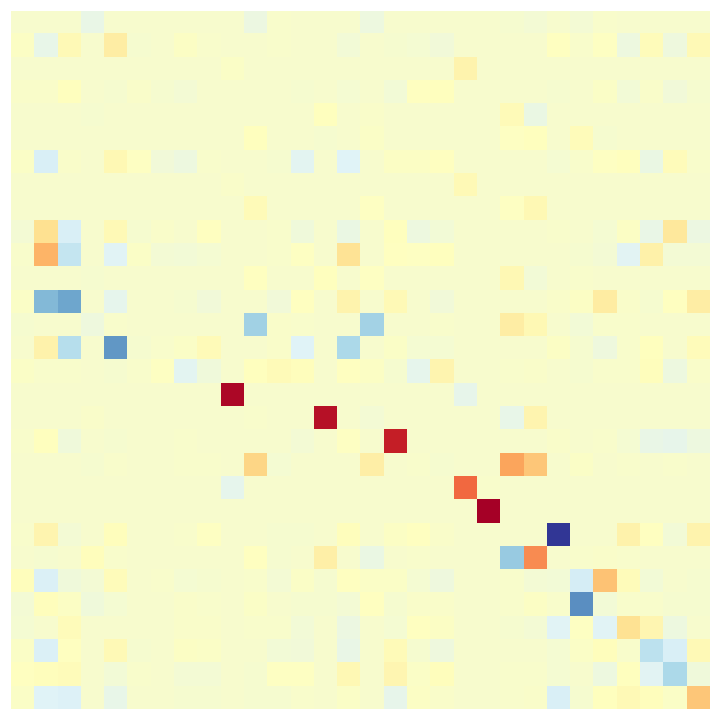}    %
		\end{tabular}
		\caption{\label{fig:bandpass}
			Band-pass filtering of a simplicial complex using our method. The top row shows the input and coarsened complexes. Ours is executed with two different input specifications. Rows 2 and 3 depict the functional maps of the lowest ten and highest 100 frequencies of $L_1$. Gudhi's method is specifically designed to preserve homology so its map is only diagonal for the first 4 (number of holes) elements. Our algorithm can controllably coarsen the input complex. 
		}
		\label{fig:bands}
	\end{center}
\end{figure}

\subsection{Applications}
\paragraph{Denoising}

A straightforward application of our method is to suppress frequencies associated with noise. We filtered a noisy \texttt{Bunny} model (43K vertices) using a very narrow low-pass filter: only the first three eigenvectors of $L_0$ and $L_1$.  Figure~\ref{fig:denoise} visualizes our result along with the baseline filters unevenly, possibly due to noisy geometric information in its cotan weighting.

\newcommand{\bunnySURF}{0.1621}
\begin{figure}
	\begin{center}
		\begin{tabular}{@{}c@{}c@{}c@{}}
			noisy & ours& baseline \\		
			\includegraphics[width=\bunnySURF\textwidth]{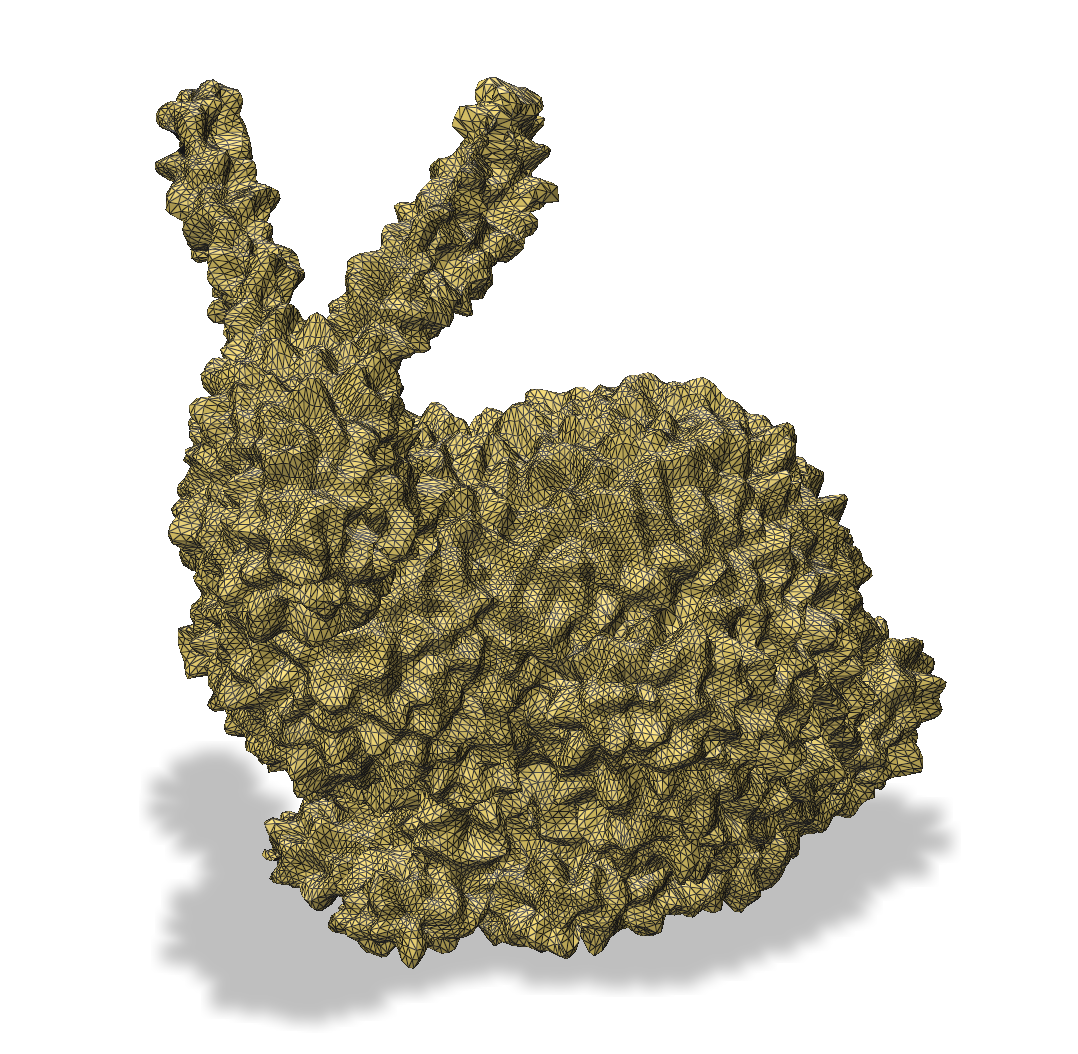} &			
			\includegraphics[width=\bunnySURF\textwidth]{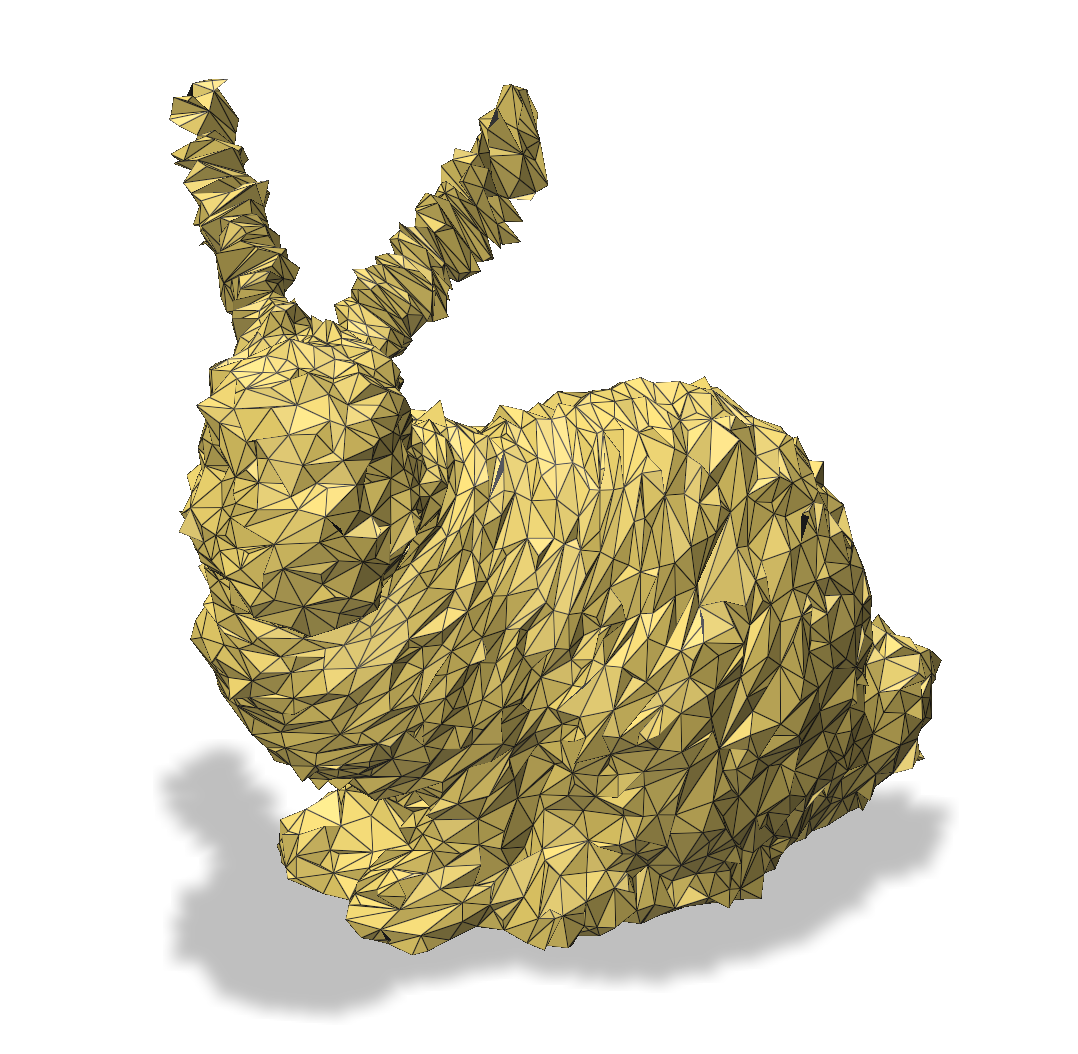} &			
			\includegraphics[width=\bunnySURF\textwidth]{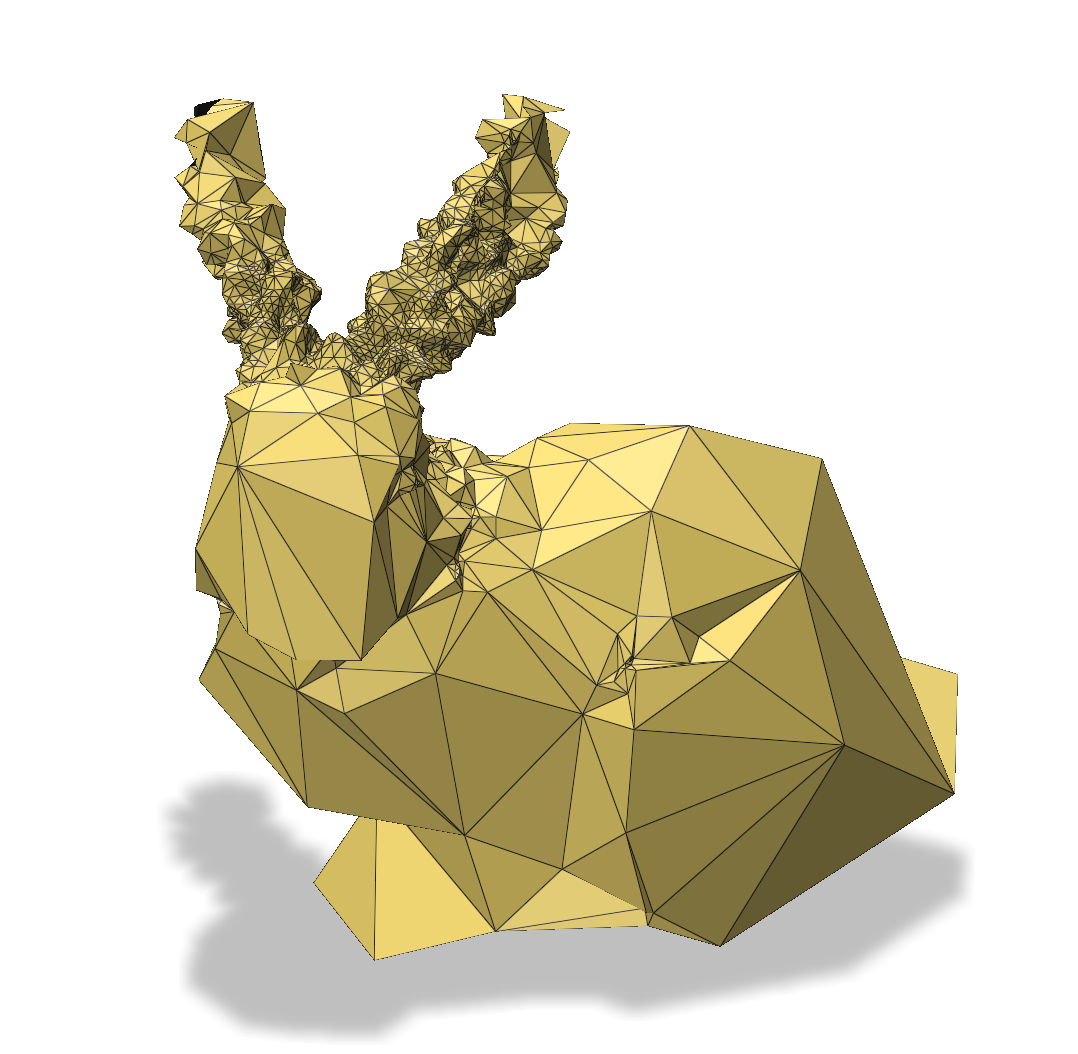}
		\end{tabular}
	\end{center}
	\caption{\label{fig:denoise} Denoising. A noisy \emph{Bunny} model (43645 vertices) is coarsened to 2\% of its size (1000 vertices), to preserve only the first 3 eigenvectors of $L_0$ \& $L_1$ using our method, and the first 3 eigenvectors of $L_0^\text{cot}$ for the baseline.} 
\end{figure}

\begin{figure} [b!]
	\begin{center}
		\includegraphics[width=.99\linewidth]{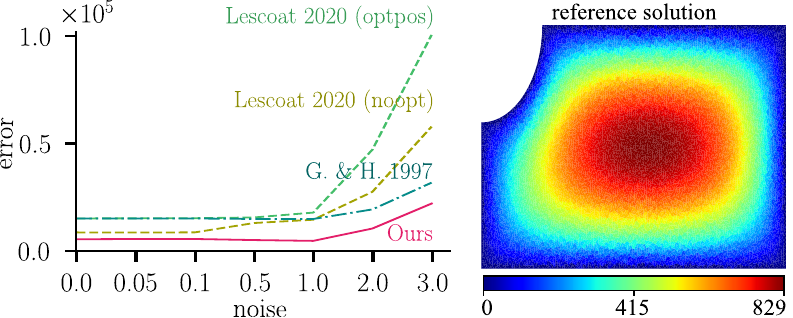} \\
		\vspace{1em}
		\begin{tabular}{@{}c@{\,}c@{\,}c@{\,}c@{}}			
			\multicolumn{4}{c}{Error maps} \\
			
			\rotatebox{90}{Ours}&
			\includegraphics[width=.315\linewidth]{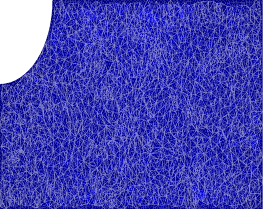} &		
			\includegraphics[width=.315\linewidth]{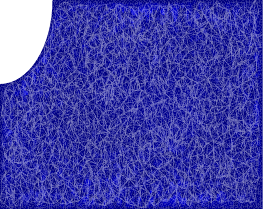} &
			\includegraphics[width=.315\linewidth]{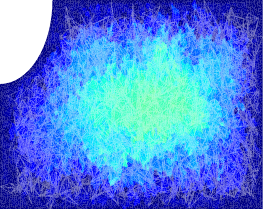} \\
			\rotatebox{90}{L. 2020 (noopt)}&
			\includegraphics[width=.315\linewidth]{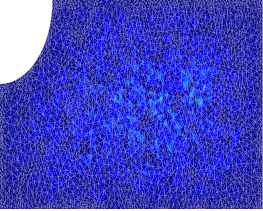} &		
			\includegraphics[width=.315\linewidth]{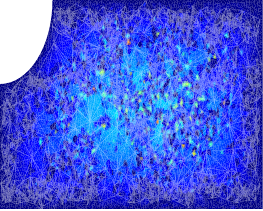} &
			\includegraphics[width=.315\linewidth]{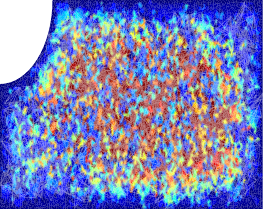} \\
			\rotatebox{90}{G\& H 97}&
			\includegraphics[width=.315\linewidth]{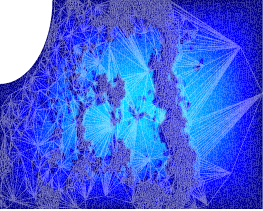} &
			\includegraphics[width=.315\linewidth]{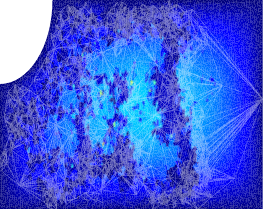} &		
			\includegraphics[width=.315\linewidth]{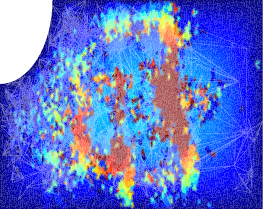} \\
			&noise = 0.0 & noise = 1 & noise = 3 
		\end{tabular}	
		\caption{\label{fig:FEM} FEM on our coarsened mesh is robust to Gaussian noise added to the vertices (top left). The solution to $-\Delta u = 1$ on the fine mesh (zoom in to view) is shown on the top right. The table below compares error as heat maps of simulation error (blue is low and red is high), with the coarse mesh (zoom to view) overlaid, for ours and two other coarsening methods (rows). }
	\end{center}
\end{figure}

\paragraph{Finite Element Method (FEM)} We solve the Poisson equation $-\Delta u = 1$ with Dirichlet boundary condition $u=0$ on the popular \texttt{Plate-hole} model, using piece-wise linear (triangular) elements. In addition to solving it on the discretized mesh, we test robustness by applying planar pertubations to the vertices with increasing levels of Gaussian noise. For each setting, we coarsen the noisy mesh with $\rho=0.75$ (8000 to 2000 vertices)  using $L_0^\text{cot}$ and $L^1$, run a standard FEM solver on the coarse mesh and lift the solution to the input mesh. Figure~\ref{fig:FEM} shows a plot of error vs noise (top left), the reference solution (top right) and error maps for different noise settings (columns) and methods (rows). Also shown are previous work~\cite{lescoat2020spectral}, with (optpos) and without (noopt) vertex position optimization, and a quadric-based method~\cite{garland1997surface}. The coarse mesh is overlaid on the error maps and can be viewed by magnifiying the figure. Our method exhibits robustness against noise and improved approximation performance against element reduction.

\paragraph{Spectral distances} We evaluate the fidelity (Figure~\ref{fig:dists}) of spectral distance measures computed on the coarse mesh and ``lifted'' to the fine mesh, between vertices $w,v$ of a triangle surface mesh (\texttt{Suzanne}) and a volume tetrahedral mesh (\texttt{Fertility}), using the same parameters as Figures~\ref{fig:eigenvectors2D} and~\ref{fig:eigenvectors3D}. Figure~\ref{fig:teaser} illustrates the spectral distance approximation process on the \emph{Engine} tetrahedral model, coarsened with $\rho=-.8$ (from 46220 to 10000 vertices) while preserving the first 50 eigenvectors of $L_0^\text{cot}$ and the first 25 eigenvectors of $L_1$ and $L_2$. The model has 20 holes that are retained in the coarse mesh. The metrics used and their parameters are tabulated below. 

\begingroup
\renewcommand{\arraystretch}{1.3} 
\begin{tabular}{lrl}
\hline
     diffusion&$w,v,t$&$\sum_i  (u_i(v)-u_i(w))^2e^{-2s_it}$ \\
     biharmonic&$w,v$&$\sum_i (u_i(v)-u_i(w))^2/s_i^2$ \\
     commute&$w,v$&$\sum_i (u_i(v)-u_i(w))^2/s_i$ \\
     WKS&$v,t$&$\sum_i u_i^2(v)e^{-\frac{(t-\log s_i)^2}{2\sigma^2}}/\sum_i e^{-\frac{(t-\log s_i)^2}{2\sigma^2}}$ \\
     WKD & $w,v$&$\int_{t_\text{min}}^{t_\text{max}} \left|\frac{WKS(w,t)-WKS(v,t)}{WKS(w,t)-WKS(v,t)} \right|dt$, \\
     HKS&$w,v,t$&$\sum_i u_i(w)u_i(v)e^{-s_it}$ \\
\hline
\end{tabular}
\endgroup

Spectral distance approximation on \texttt{Suzanne} outperforms the baseline by orders on magnitude in most cases, with error remaining low even for the tetrahedral \texttt{Fertility} mesh. The qualitative comparison indicates agreement between our ``lifted" version and the reference, despite some localized distortions.

\section{Discussion}

\begin{figure} [htbp]
	\begin{center}
		\begin{tabular}{@{}c@{\,}c@{\,}}			
			\includegraphics[width=.49\linewidth]{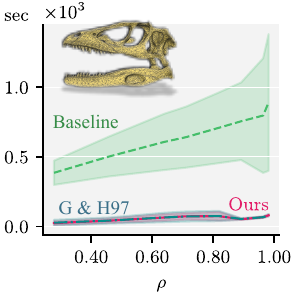} &		
			\includegraphics[width=.49\linewidth]{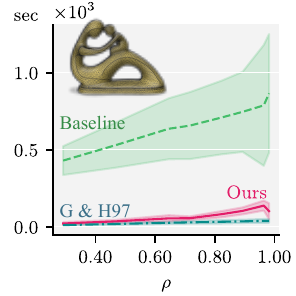} 
		\end{tabular}	
		\caption{\label{fig:timing} Computation time vs reduction ratio. }
	\end{center}
\end{figure}

\paragraph{Execution time} We implemented our method in C++, and performed experiments on a 16-core workstation (Intel E5-2630 v3,2.4 GHz) with 64GB RAM. Our method compares favorably (orders of magnitude faster) against the spectral coarsening baseline in Figure~\ref{fig:timing}. Although it appears that we are competitive with the quadric-based method~\cite{garland1997surface}) it should be noted that ours (like the baseline) requires an eigenspace of the laplacians as input while the latter does not. The time for the eigendecomposition is not reflected in the plot (which only measures geometric operations) and should be added to both our method and the baseline. The triangle meshes used in the comparison contain 10772 (\texttt{Fertility}) and 29690 (\texttt{Dinoskull}) vertices, respectively. We use $L_0^\text{cot}$ for the baseline, and $L_0^\text{cot}$ \& $L_1$ for our method.

\paragraph{Boundary} The intricacies of boundary values problems involving Hodge Laplacians~\cite{mitreaHodgeLaplacianBoundaryValue2016} complicate practical computations. The differential operators need to be constructed carefully to guarantee coverage of tangential and normal conditions~\cite{zhao3DHodgeDecompositions2019b}. Since our method is Laplacian-agnostic, any positive semi-definite variante of the Laplacian operator, with its spectrum, can be provided as input.




\paragraph{Limitation and future work} The utility of our tool hinges on knowledge of spectral constraints imposed by downstream tasks. While this is known in some cases (homology, spectral distances and denoising), such constraints are not typically known across applications. There is especially little use of higher dimensional Laplacian operators. However, we are hopeful that the availability of a tool such as ours will inspire the graphics communities (particularly geometry processing and simulation) to explore the applicability of mixed-dimensional Laplacian spectral constraints.

\section{Conclusion} \label{sec:concl}

We presented a simple and efficient algorithm for coarsening simplicial complexes, while preserving targeted spectral subspaces across multiple dimensions. We exemplified the impact of the choices of spectral domain on applications such as denoising, FEM and approximate distance calculations on coarsened meshes. We hope that this work will pave the way towards unleashing the potential of using mixtures of Laplacians in discrete geometry processing.

\newcommand{\evecwidthSurf}{0.195}
\newcommand{\smoothreswidth}{0.27}
\newcommand{\GwidthTRI}{0.22}

\begin{figure}
	\begin{center}
		\begin{tabular}{@{}c@{}m{\smoothreswidth\textwidth}@{}m{\evecwidthSurf\textwidth}@{}}
			{} & {} & $U_i, \, i= 1,2,9,50$ 
			\\
			\rotatebox{90} {Reference} &
			\includegraphics[width=\smoothreswidth\textwidth]{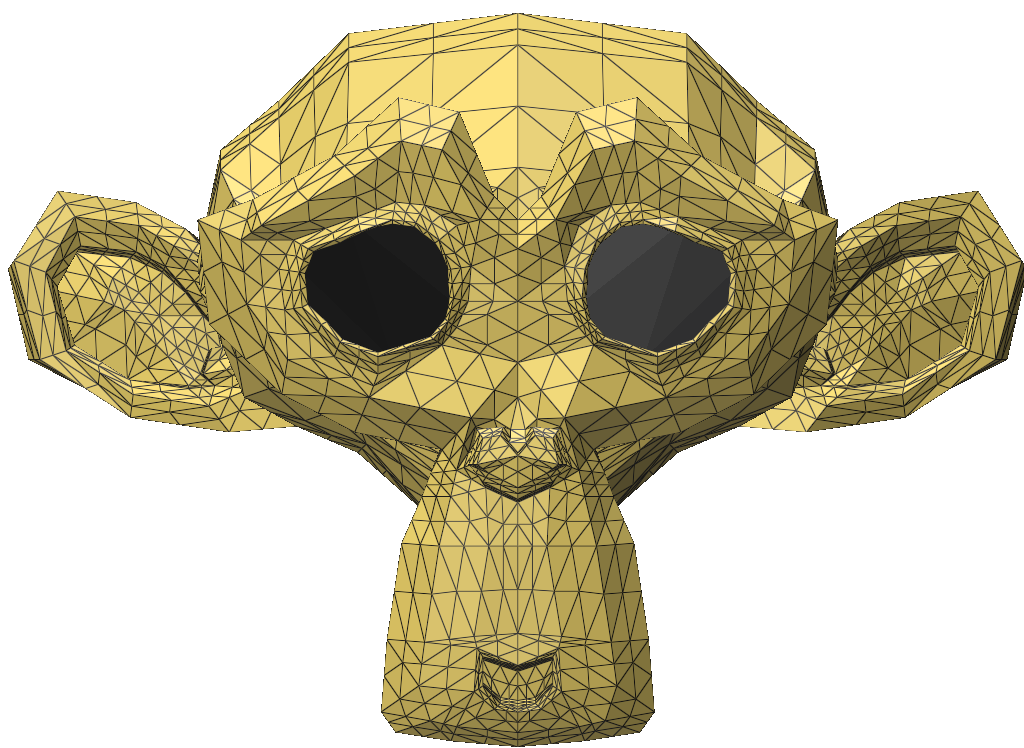} &
			\includegraphics[width=\evecwidthSurf\textwidth]{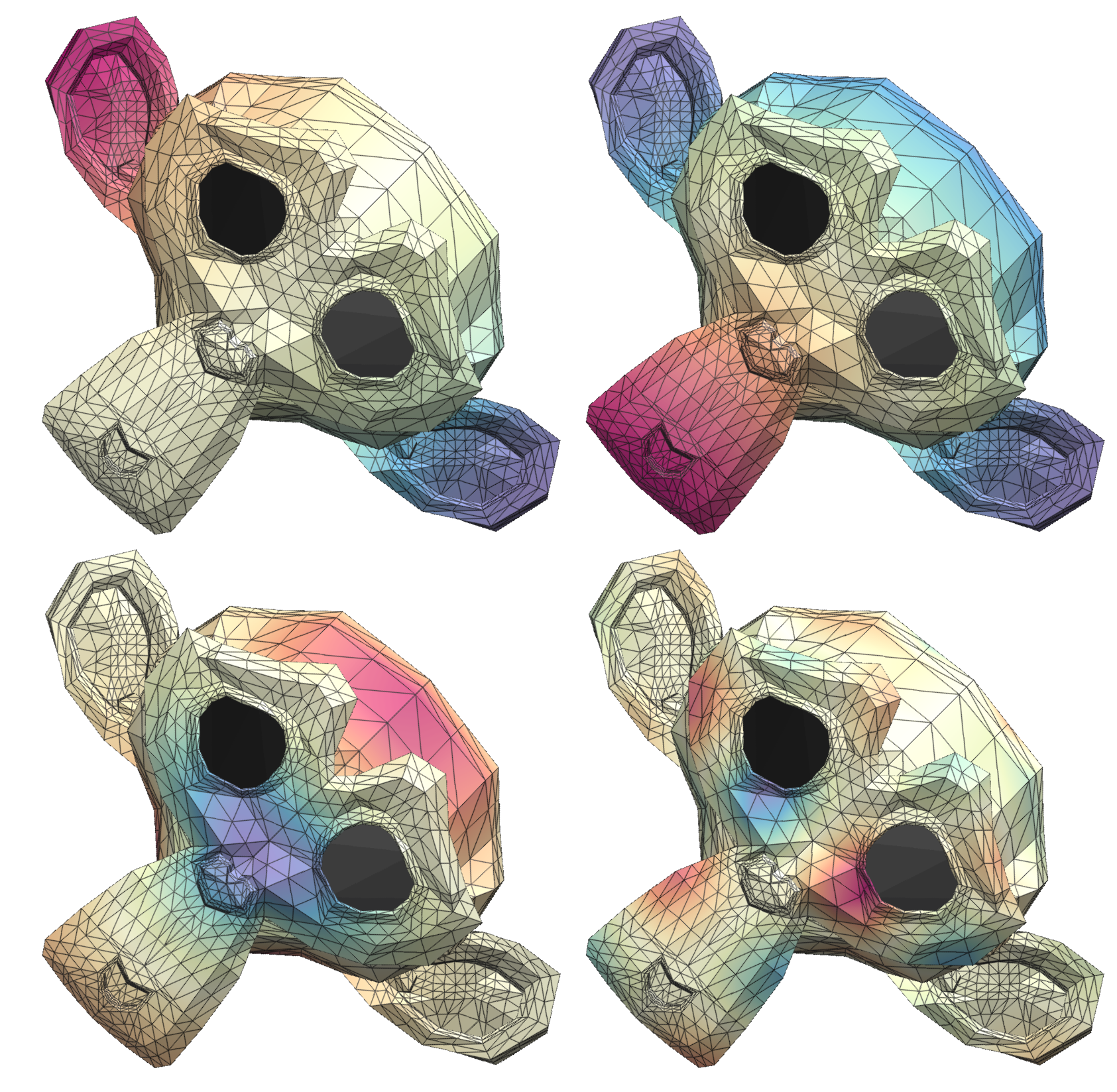} 
			\\
			\rotatebox{90} {Ours} &
			\includegraphics[width=\smoothreswidth\textwidth]{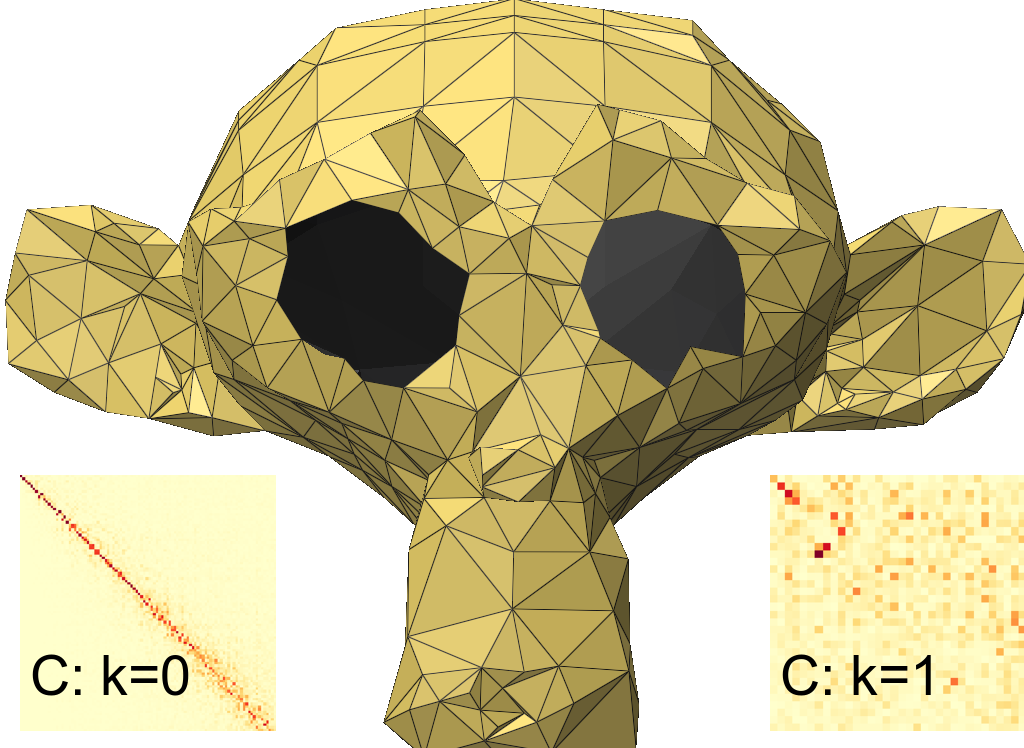} &
			\includegraphics[width=\evecwidthSurf\textwidth]{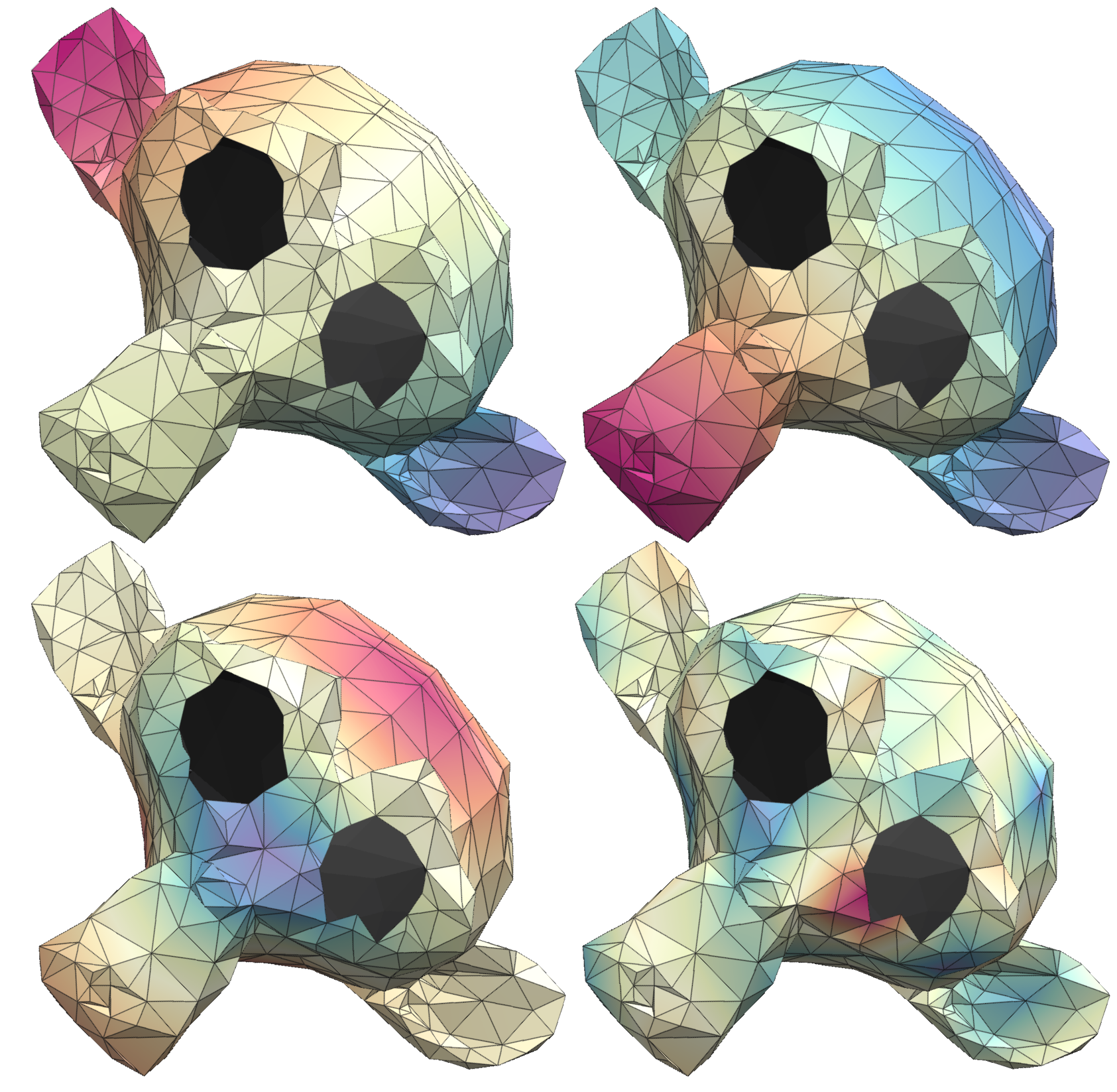} 
			\\			
			\rotatebox{90} {Baseline} &
			\includegraphics[width=\smoothreswidth\textwidth]{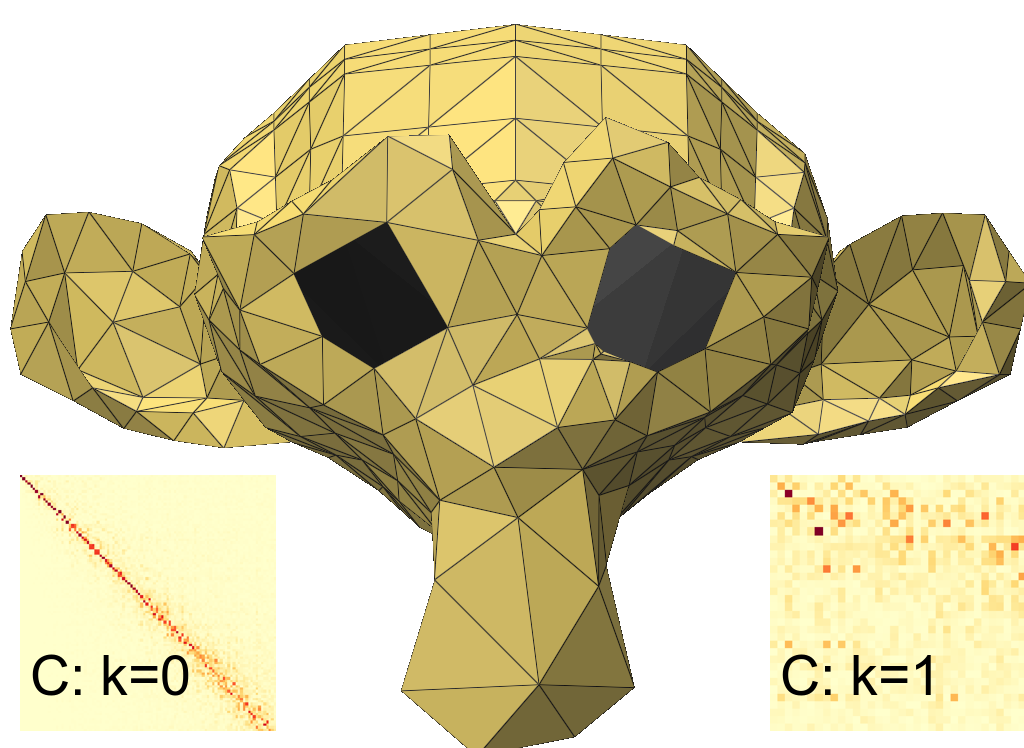} &
			\includegraphics[width=\evecwidthSurf\textwidth]{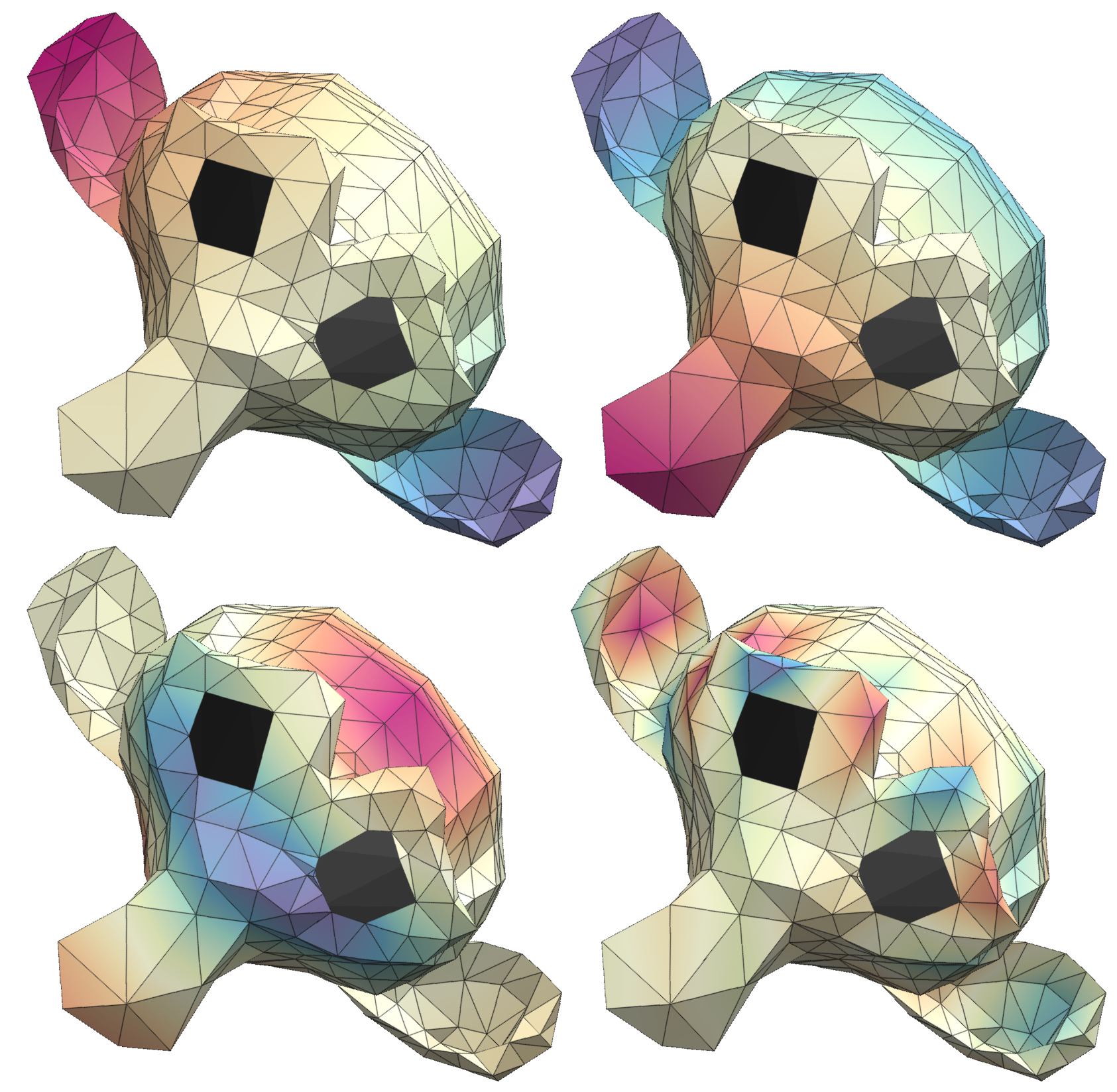} 
		\end{tabular}
		\begin{tabular}{@{}c@{}c@{}}
			\includegraphics[width=\GwidthTRI\textwidth]{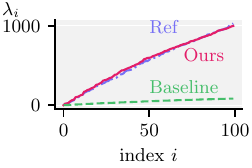} &
			\includegraphics[width=\GwidthTRI\textwidth]{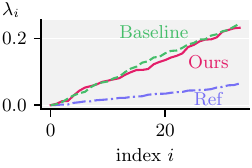}
		\end{tabular}
		\caption{\label{fig:eigenvectors2D} A visualization of  an input mesh (top left), coarsened meshes (first column) and four eigenvectors of its $L_0^{\mathrm{cot}}$ (right) computed using our method (second row) and our baseline (third row). The mesh was coarsened from 1640 to 400 vertices, preserving $L_0^{\mathrm{cot}}$ and $L_1$. Our method is mostly similar with subtle differences around the square jaw and button nose. The eigenvalues of the preserved spectrum are plotted below with those corresponding to $k=0$ on the left and $k=1$ on the right. }
	\end{center}
\end{figure}

\newcommand{\funcmapwidthTETS}{0.13}
\newcommand{\GWidthTETS}{0.142}
\begin{figure}[t!]
	\begin{center}
		\begin{tabular}{ccc}
			$L_0^\text{cot}$ & $L_1$ & $L_2$ \\
			\includegraphics[width=\funcmapwidthTETS\textwidth]{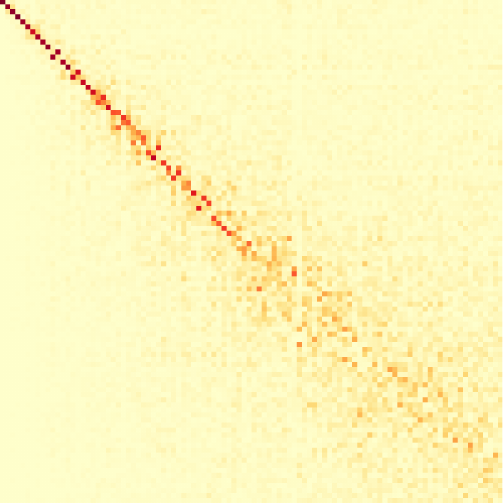} &
			\includegraphics[width=\funcmapwidthTETS\textwidth]{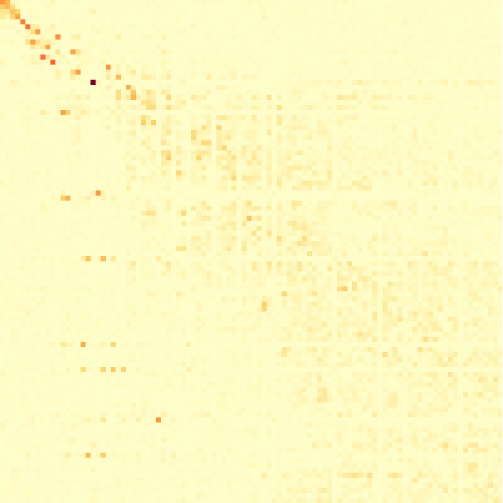} &
			\includegraphics[width=\funcmapwidthTETS\textwidth]{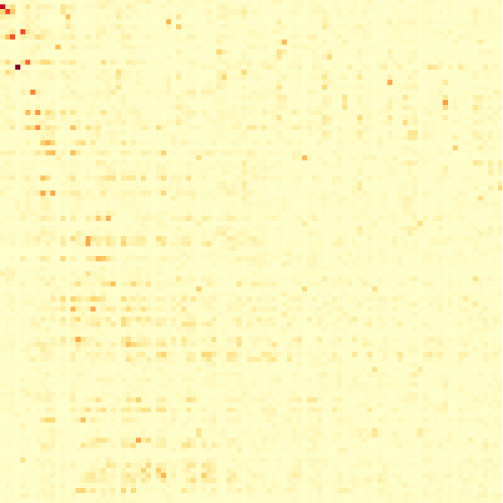} \\
			\includegraphics[width=\GWidthTETS\textwidth]{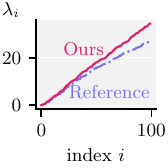} &
			\includegraphics[width=\GWidthTETS\textwidth]{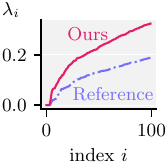} &
			\includegraphics[width=\GWidthTETS\textwidth]{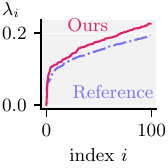}
		\end{tabular}
	\end{center}
	\caption{\label{fig:funcmaps} Functional maps (top) and eigenvalues (bottom) for the \texttt{Fertility} tetrahedral mesh pictured in Figure~\ref{fig:eigenvectors3D}.}
\end{figure}

\newcommand{\evecwidth}{0.189}
\begin{figure*}
	\begin{center}
		\begin{tabular}{@{}c@{}c@{}c@{}c@{}c@{}c@{}}
			{} & {} & $U_1$ & $U_5$ & $U_{20}$ & $U_{50}$ \\
			\rotatebox{90} {Reference} &
			\includegraphics[width=\evecwidth\textwidth]{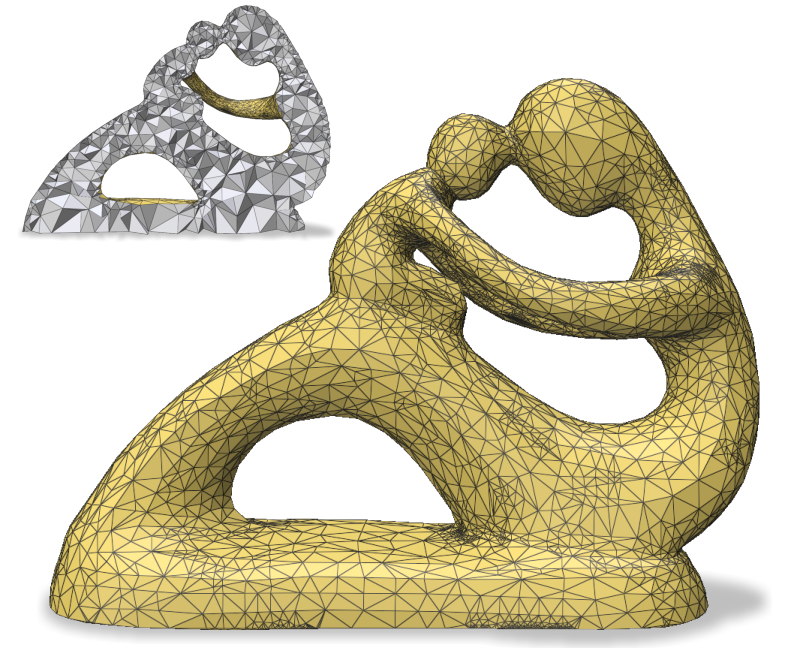} &
			\includegraphics[width=\evecwidth\textwidth]{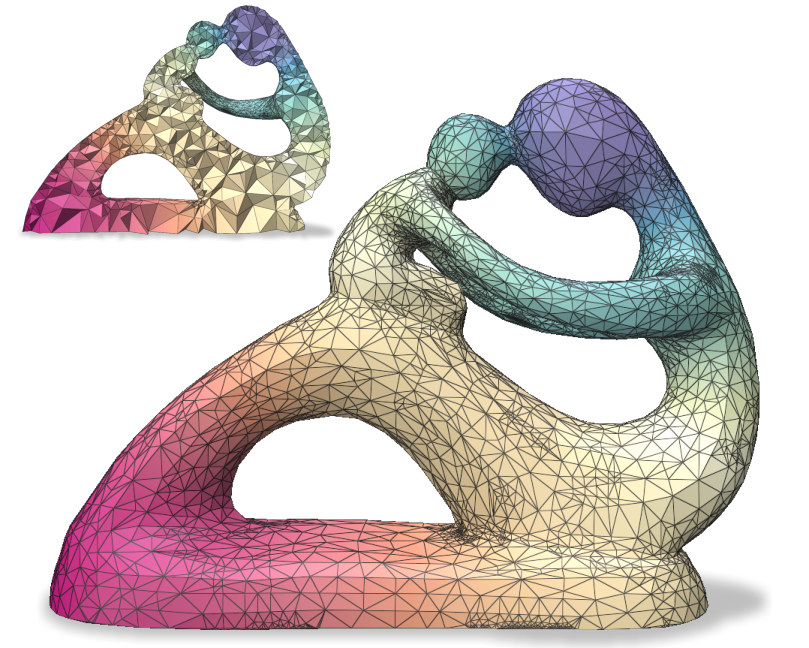} &
			\includegraphics[width=\evecwidth\textwidth]{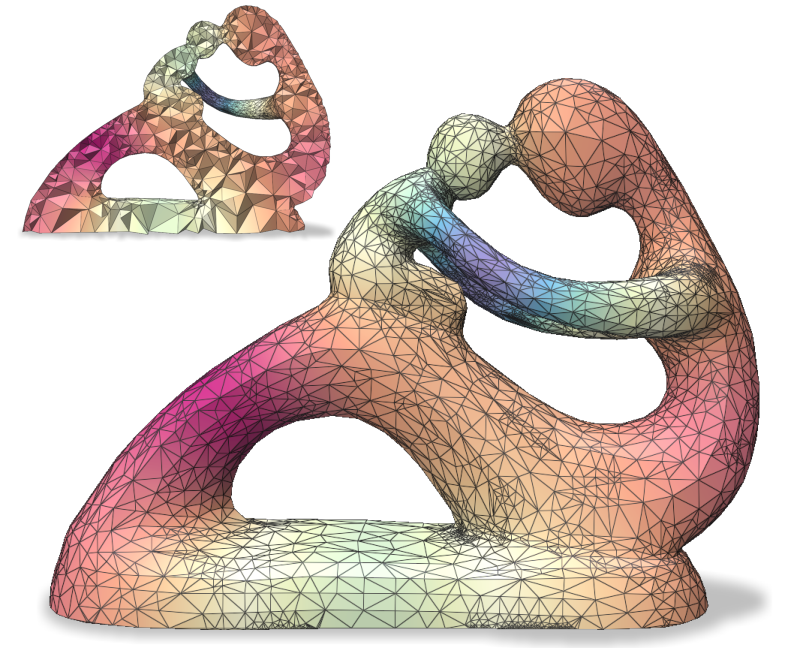} &
			\includegraphics[width=\evecwidth\textwidth]{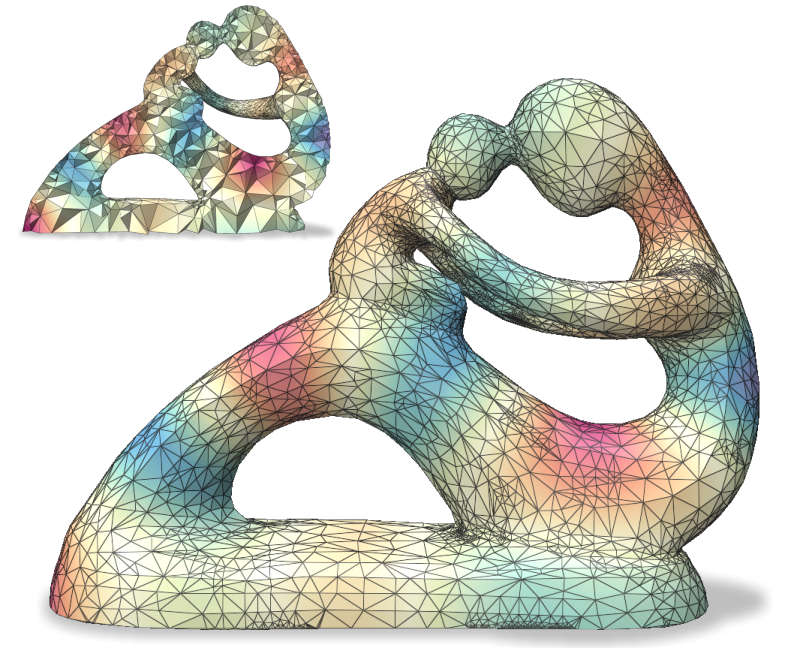} &
			\includegraphics[width=\evecwidth\textwidth]{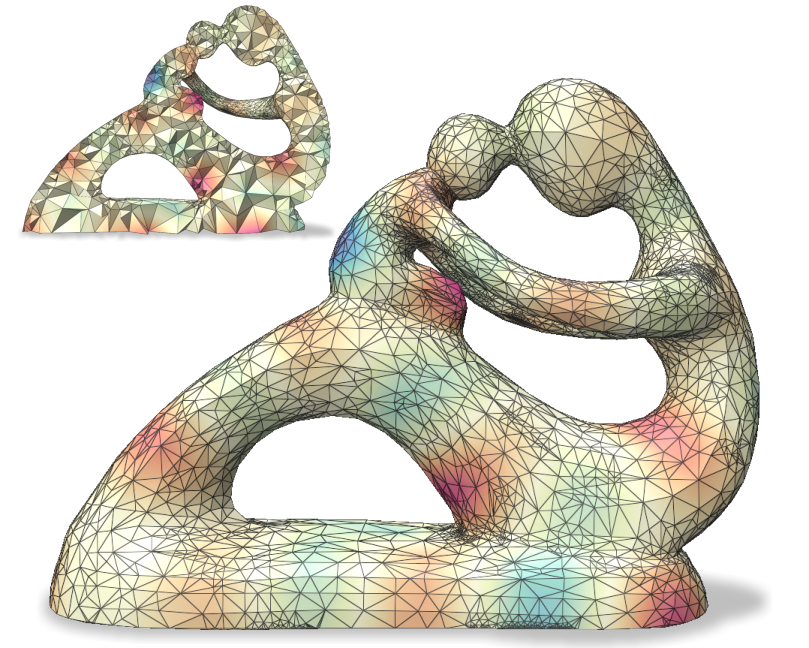} \\
			\rotatebox{90} {Ours} &
			\includegraphics[width=\evecwidth\textwidth]{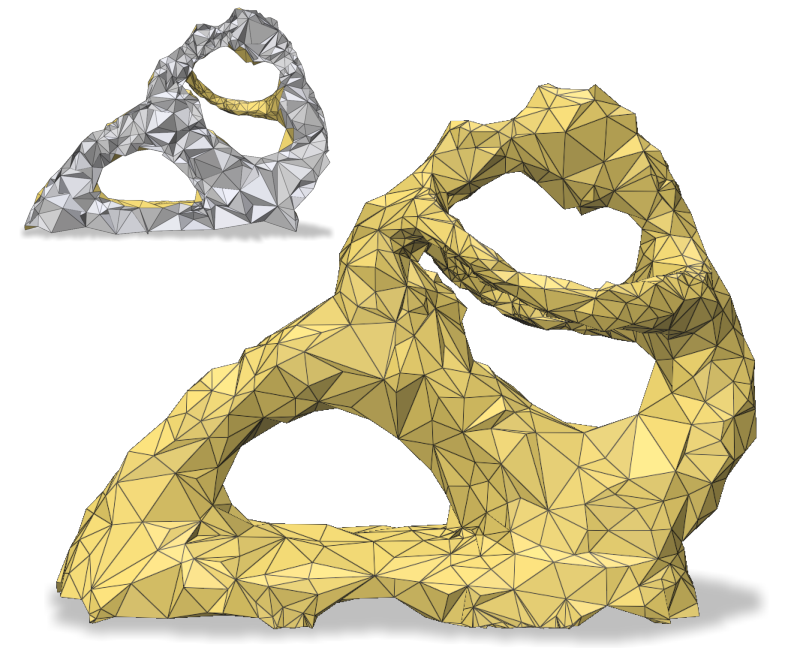} &
			\includegraphics[width=\evecwidth\textwidth]{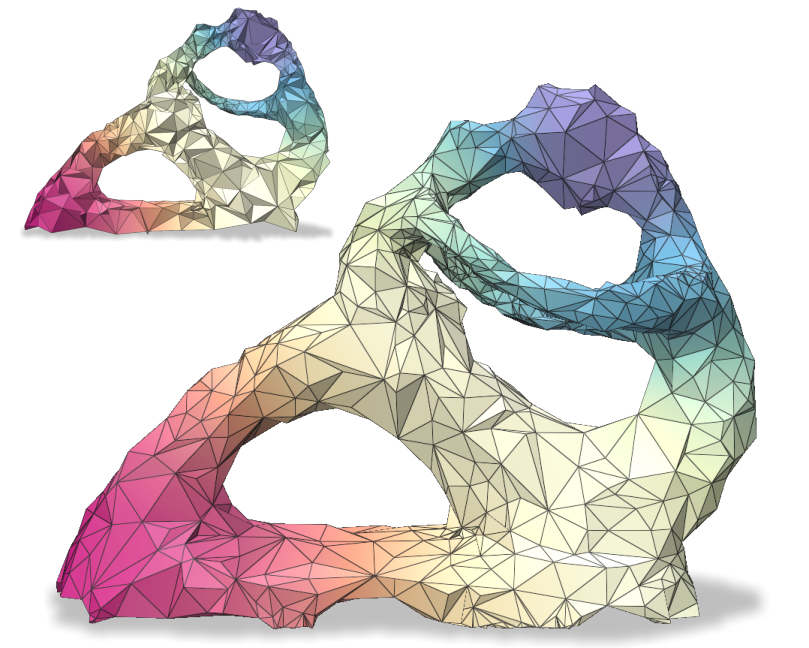} &
			\includegraphics[width=\evecwidth\textwidth]{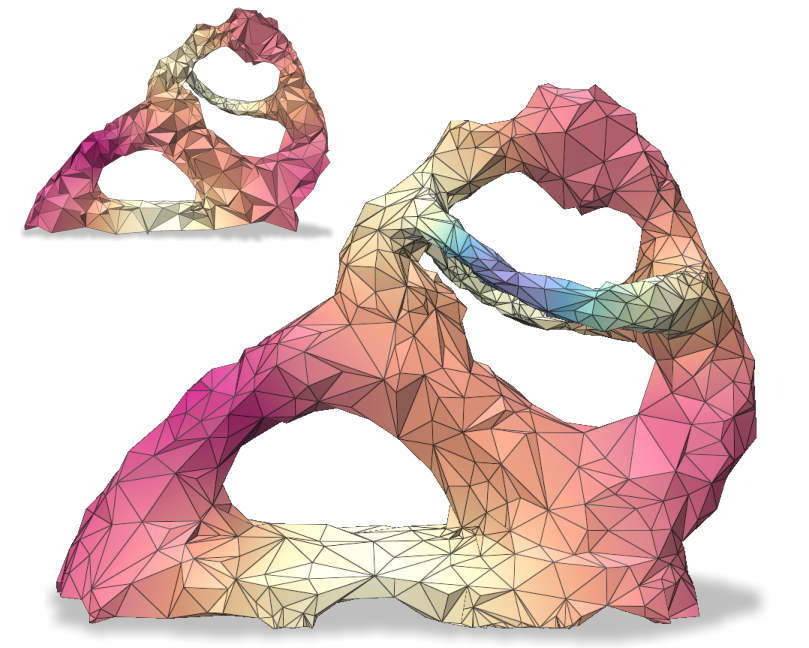} &
			\includegraphics[width=\evecwidth\textwidth]{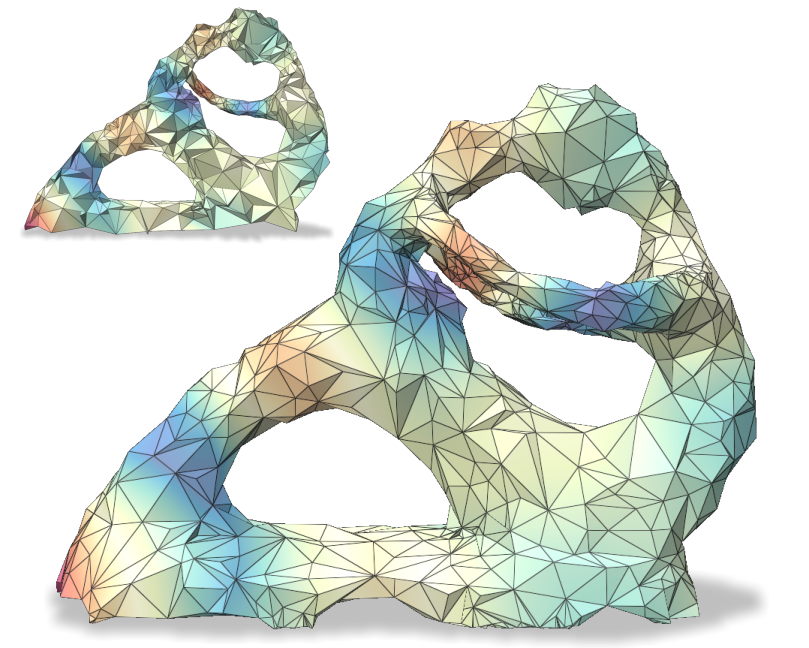} &
			\includegraphics[width=\evecwidth\textwidth]{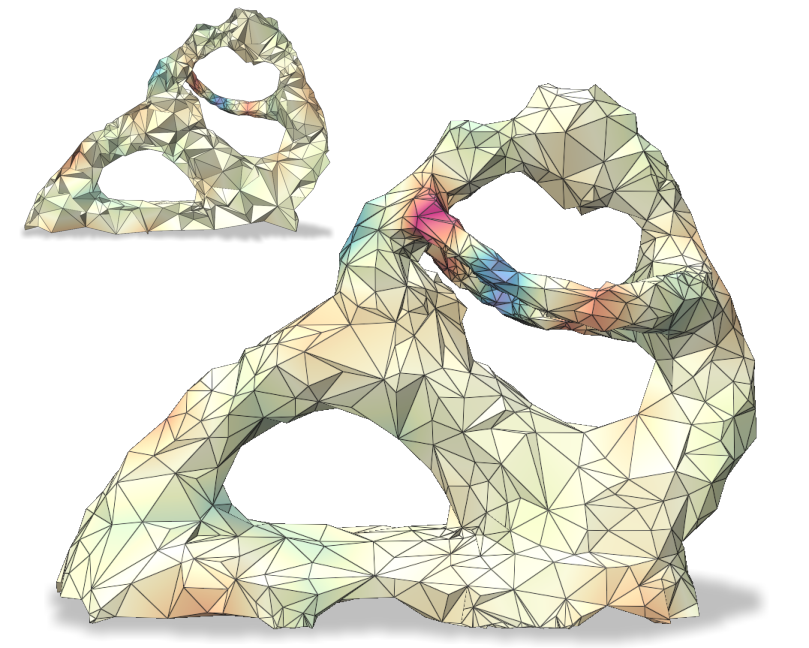}
		\end{tabular}
		\caption{\label{fig:eigenvectors3D} A visualization of four arbitrary eigenvectors $U_i$ (columns) of $L_0^{\mathrm{cot}}$ evaluated on the vertices of a tetrahedral mesh that was coarsened from 10K to 3K vertices using our algorithm. The input specification was to preserve the first 100 eigenvectors of $L_0^{\mathrm{cot}}$, $L_0$ and $L_1$. The insets on the top left depict slices through the mesh. }
	\end{center}
\end{figure*}

\newcommand{\monkeysize}{0.163}
\newcommand{\spdist}{0.32}
\newcommand{\spdistt}{0.295}
\begin{figure*}
	\begin{center}
		\begin{tabular}{@{}c@{}c@{}c@{}c@{}|c@{}c@{}c@{}c@{}}
			\multicolumn{4}{c|}{2D (triangle mesh)} & \multicolumn{4}{c}{3D (tetrahedral mesh)} \\
			\hline			
			{} & Diffusion & Biharmonic & HKS & {} & Diffusion & Biharmonic & HKS  \\
			\rotatebox{90} {Reference} & 
			\includegraphics[width=\monkeysize\textwidth]{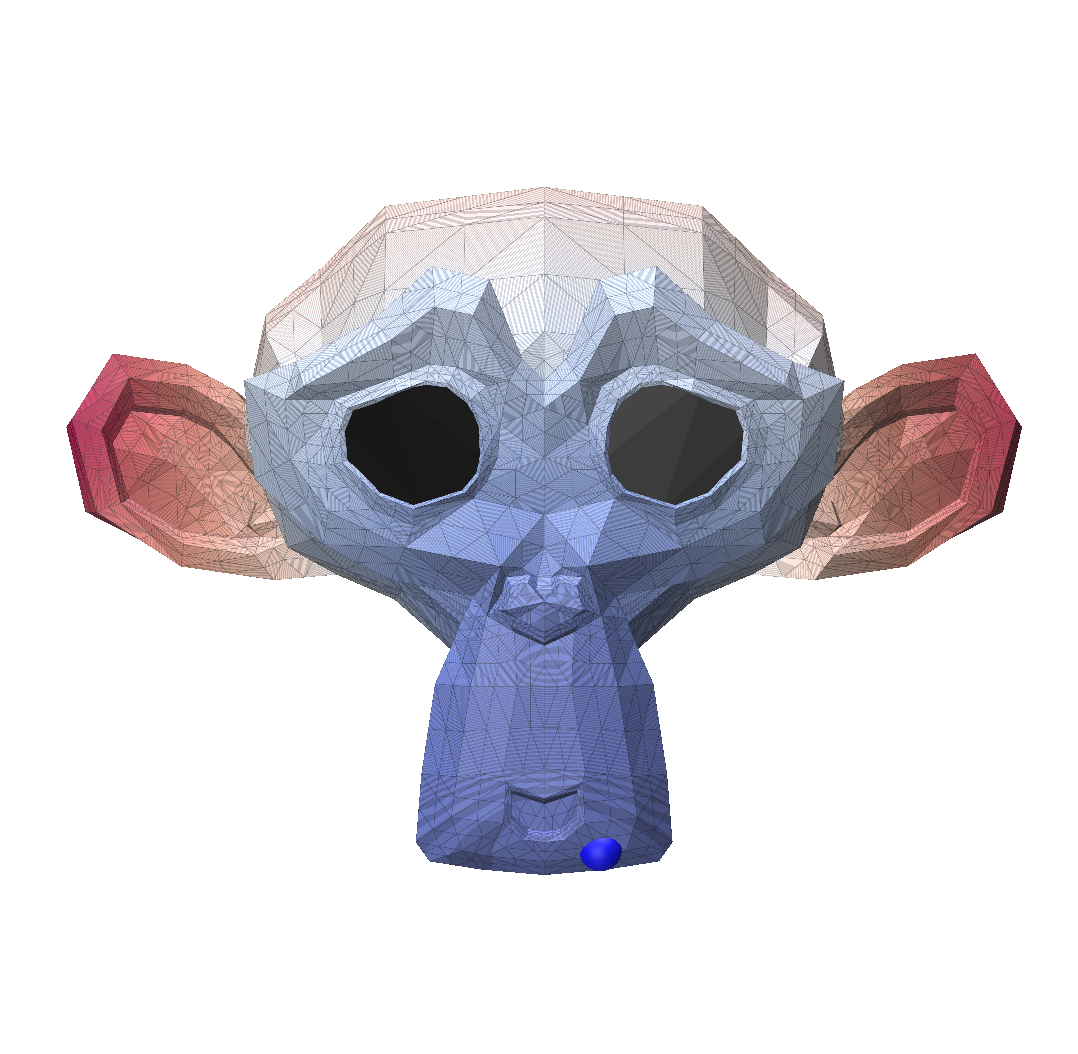} &
			\includegraphics[width=\monkeysize\textwidth]{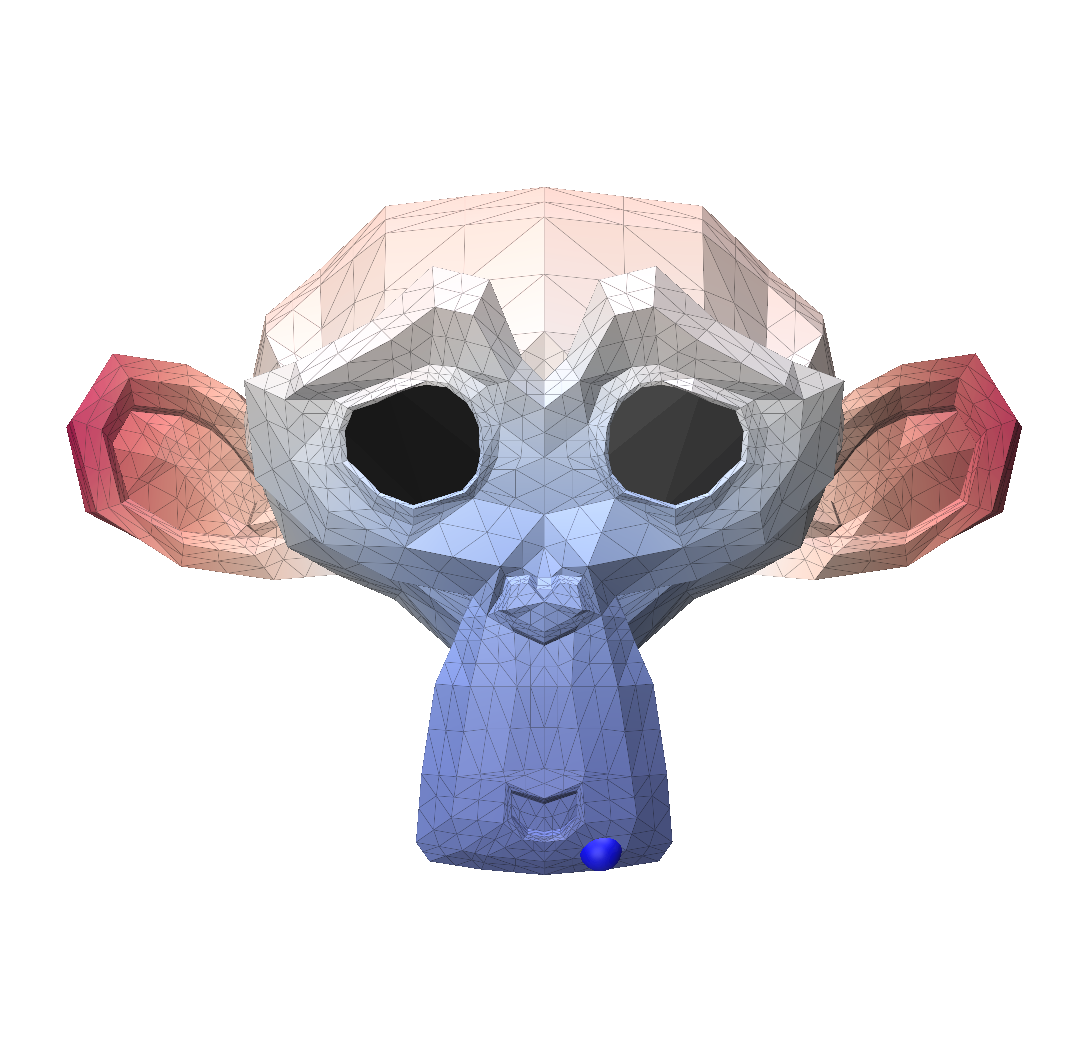} &
			\includegraphics[width=\monkeysize\textwidth]{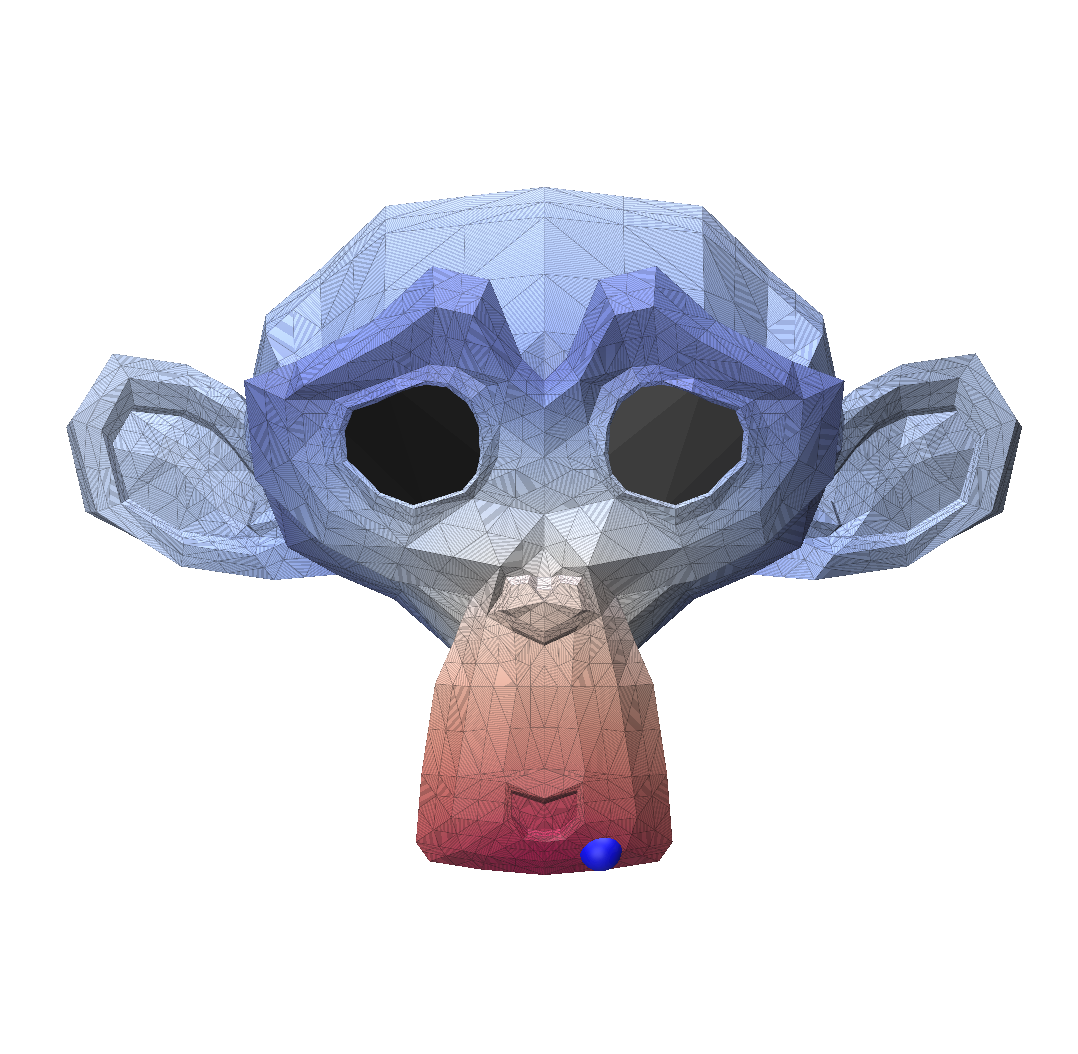}  &
			\rotatebox{90} {Reference} & 		
			\includegraphics[width=\spdistt\columnwidth]{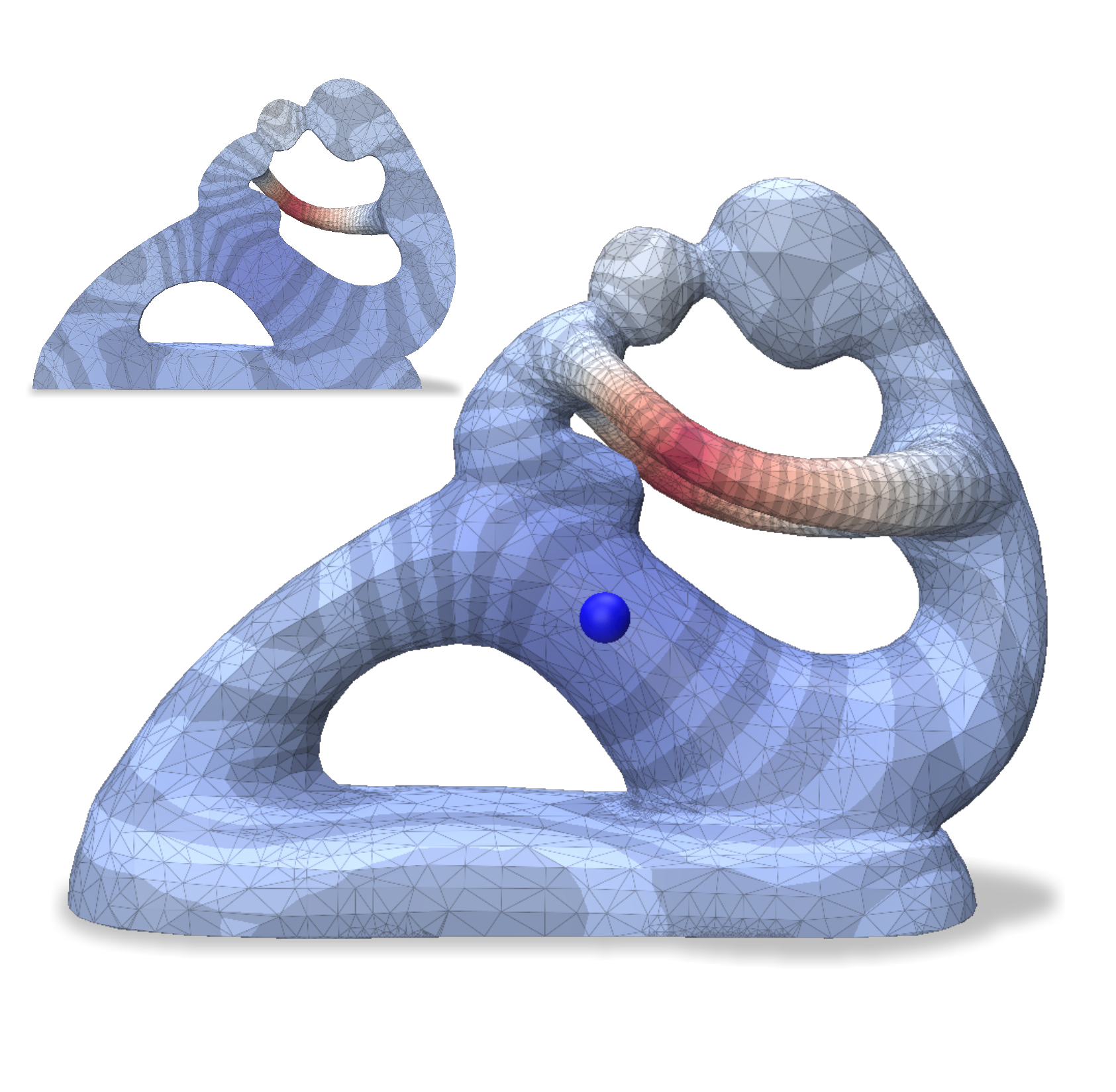} &
			\includegraphics[width=\spdistt\columnwidth]{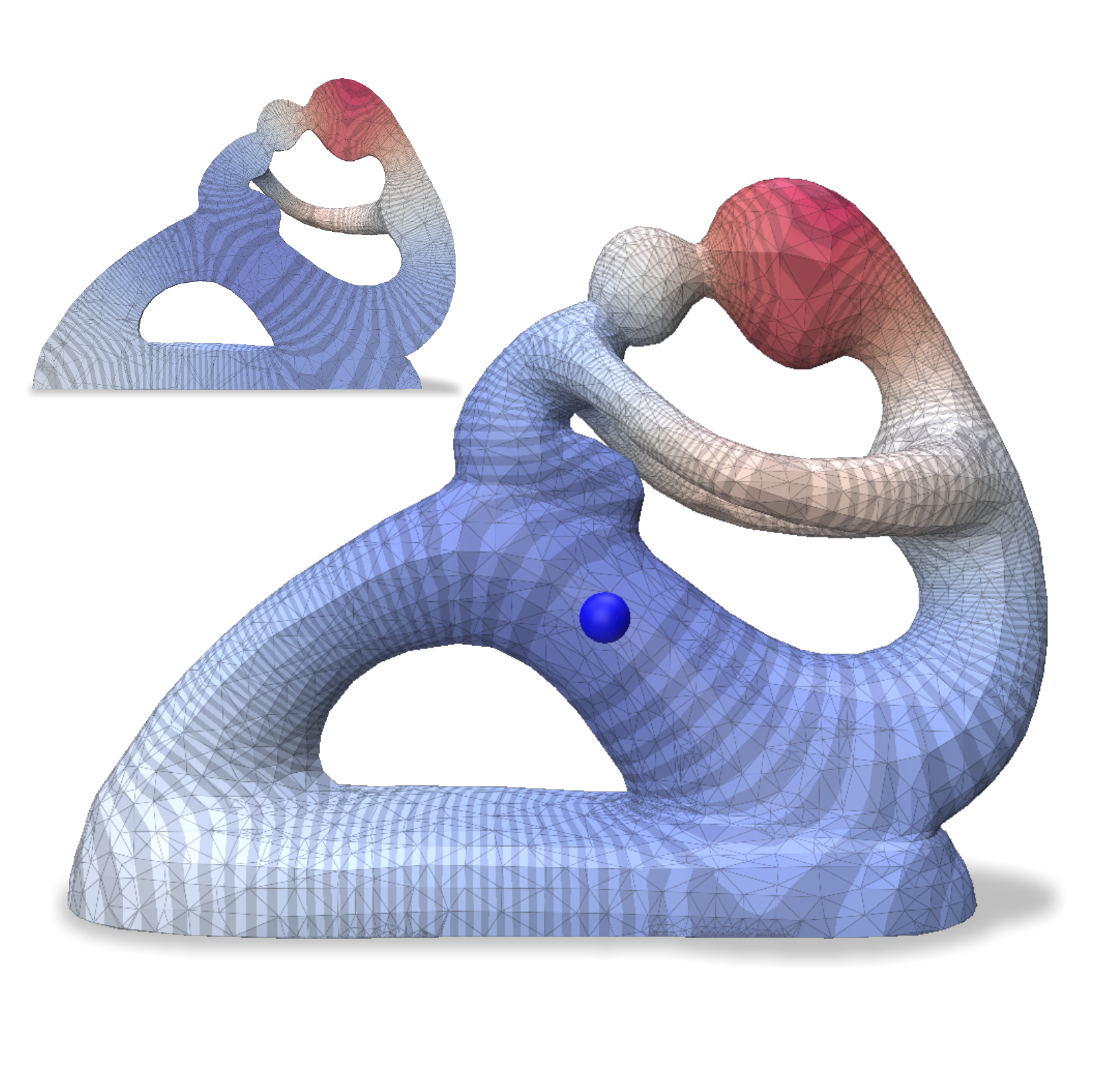} &
			\includegraphics[width=\spdistt\columnwidth]{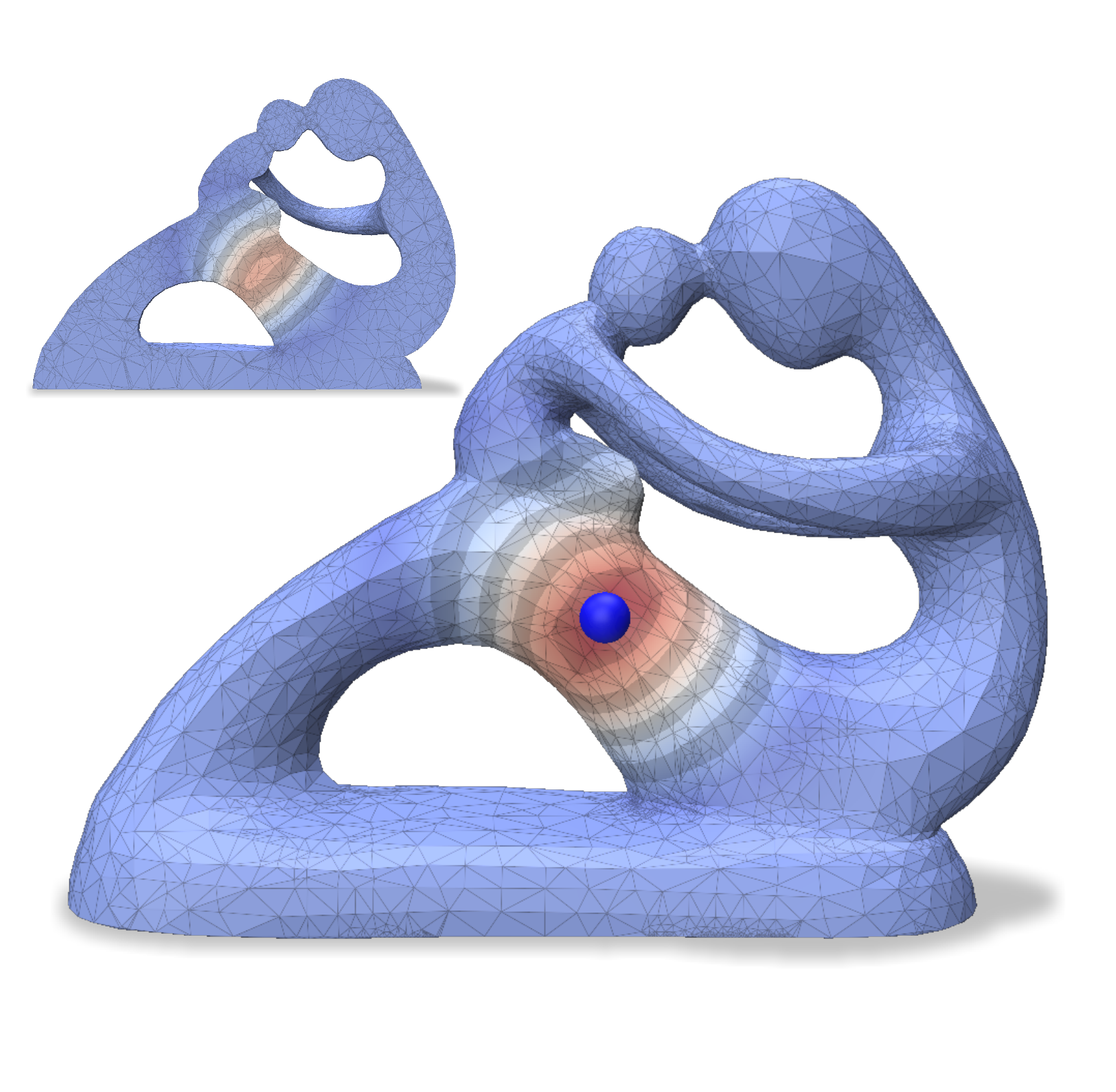} 
			\\				
			\rotatebox{90} {Ours} &
			\includegraphics[width=\monkeysize\textwidth]{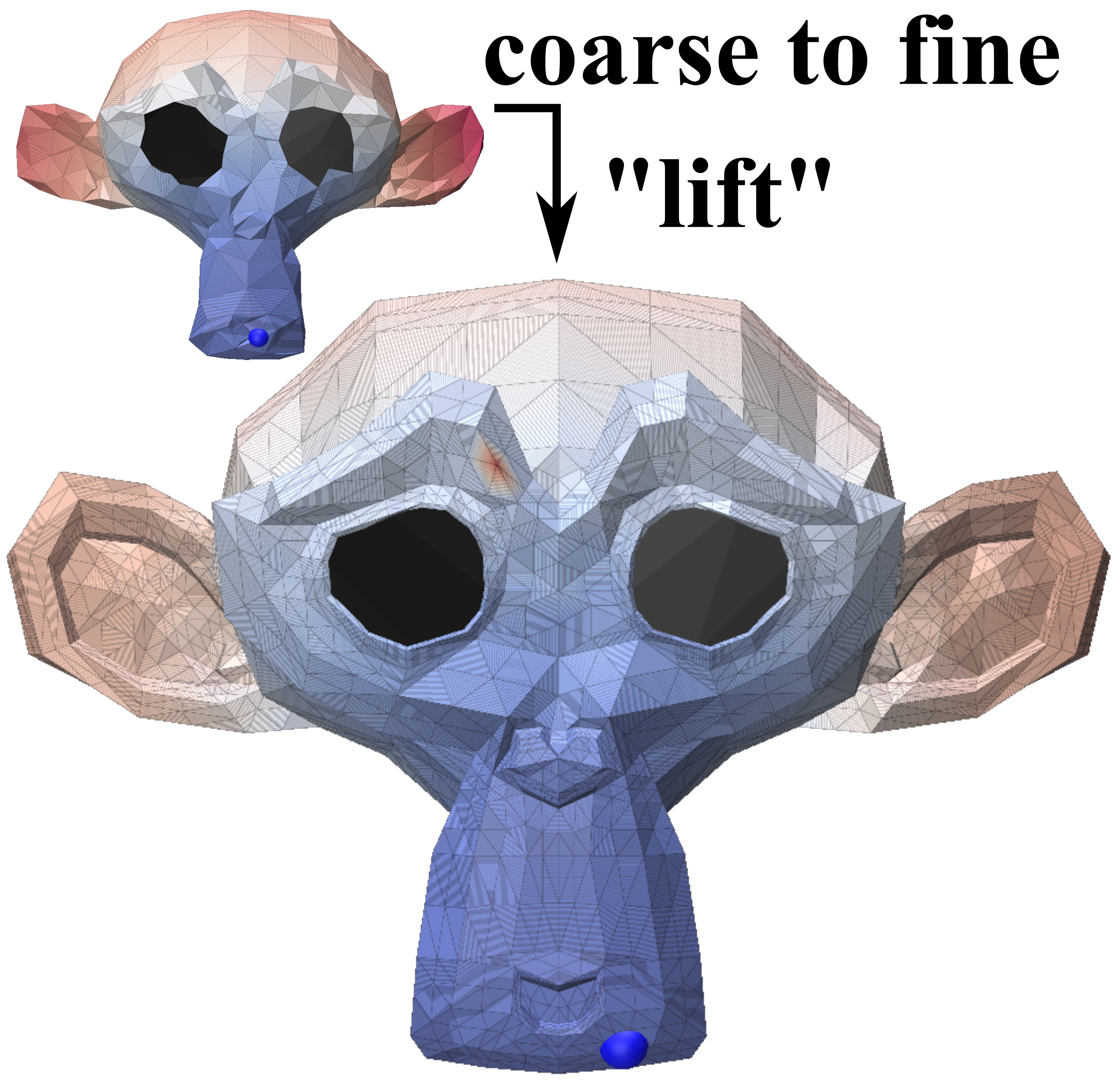} &
			\includegraphics[width=\monkeysize\textwidth]{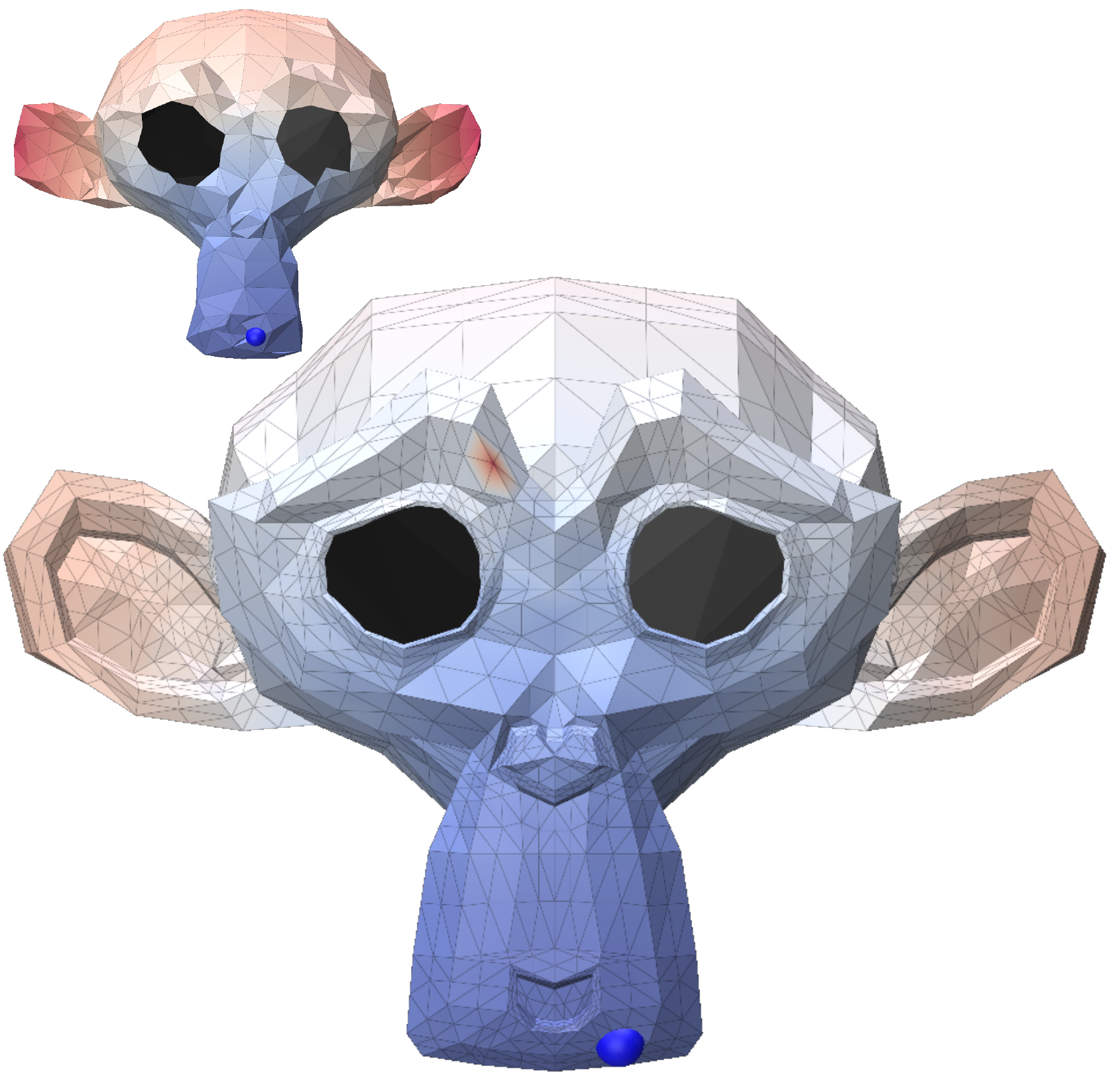} &
			\includegraphics[width=\monkeysize\textwidth]{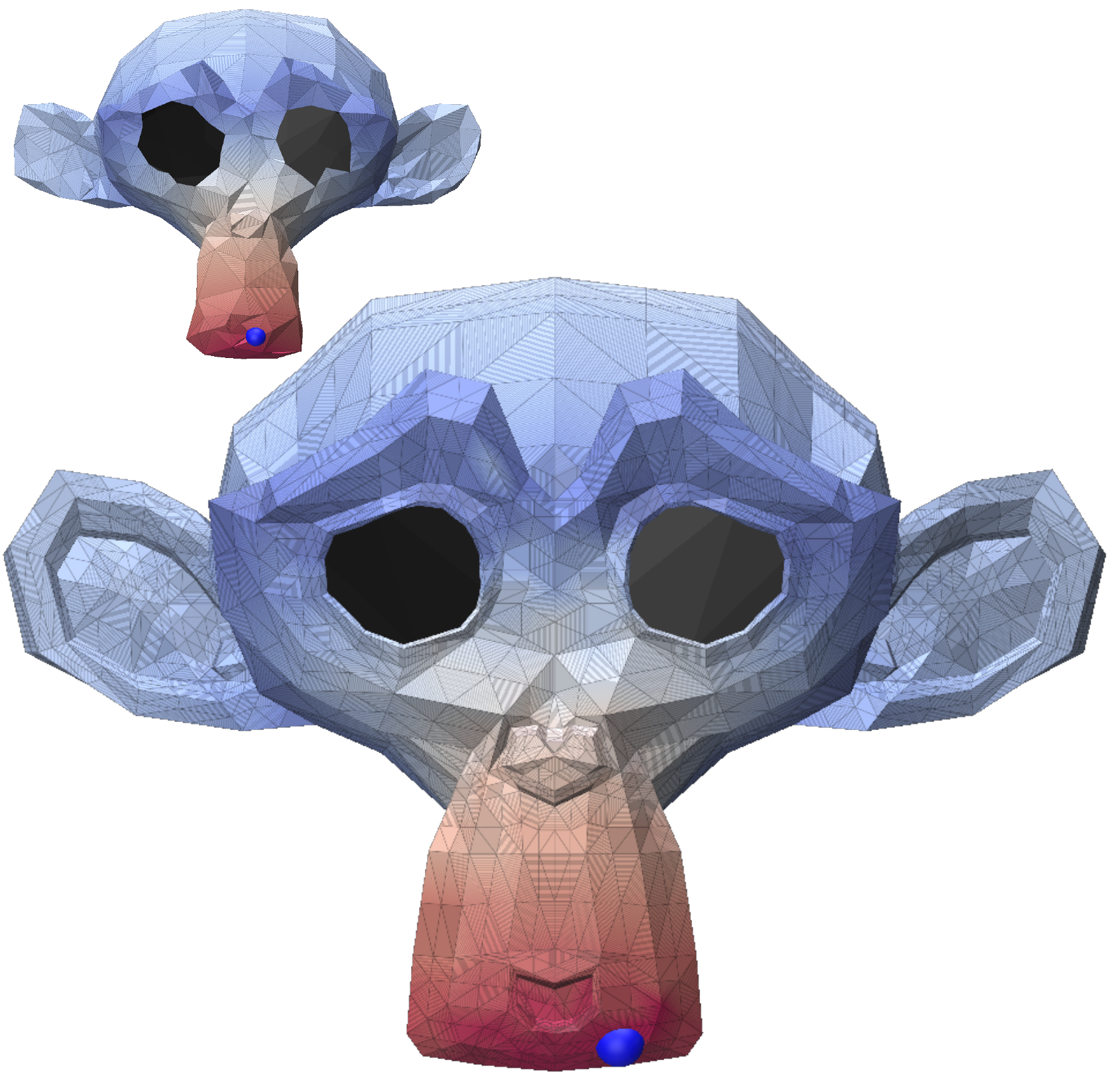} &
			\rotatebox{90} {Ours} &	
			\includegraphics[width=\spdist\columnwidth]{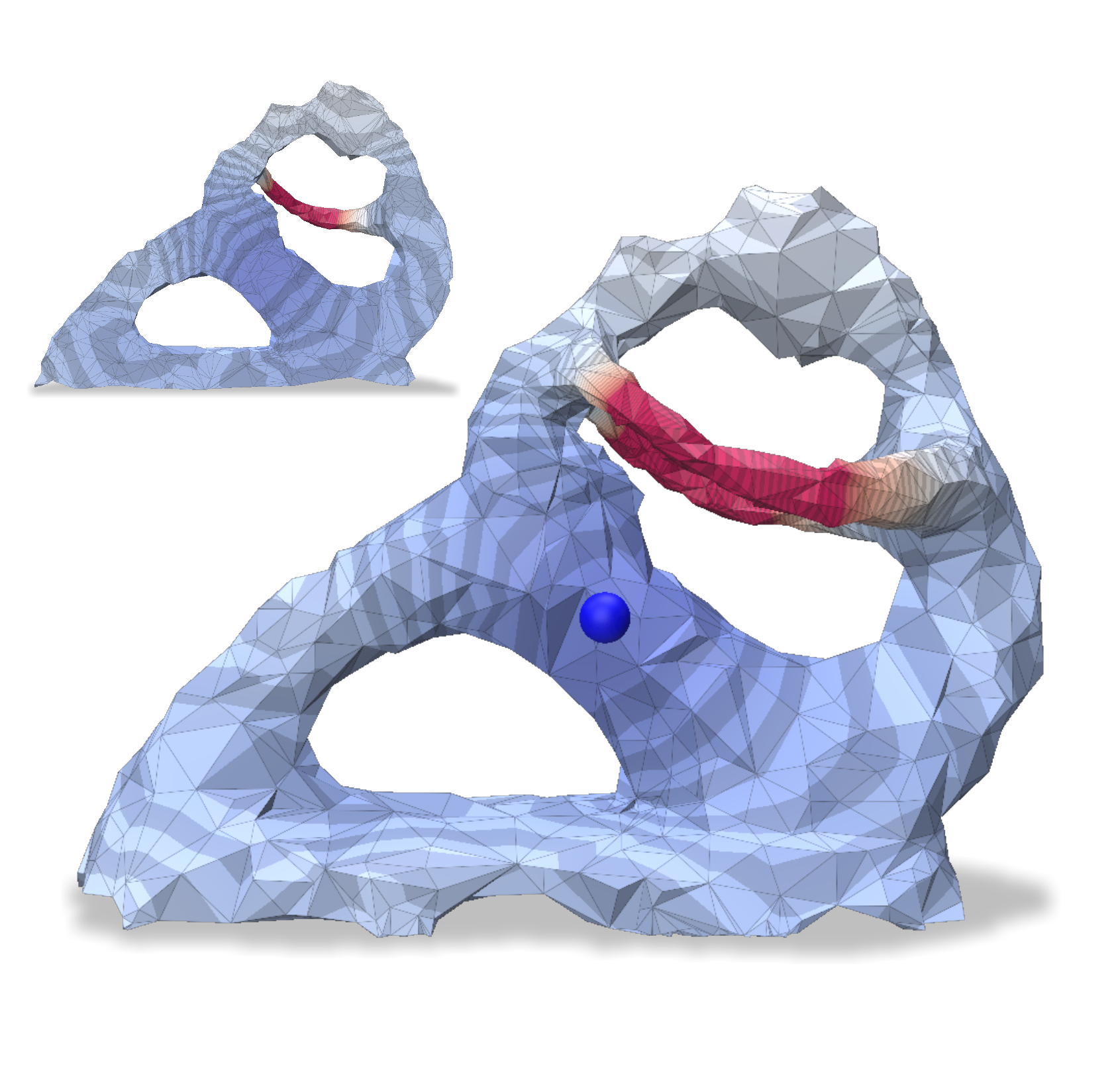} &
			\includegraphics[width=\spdist\columnwidth]{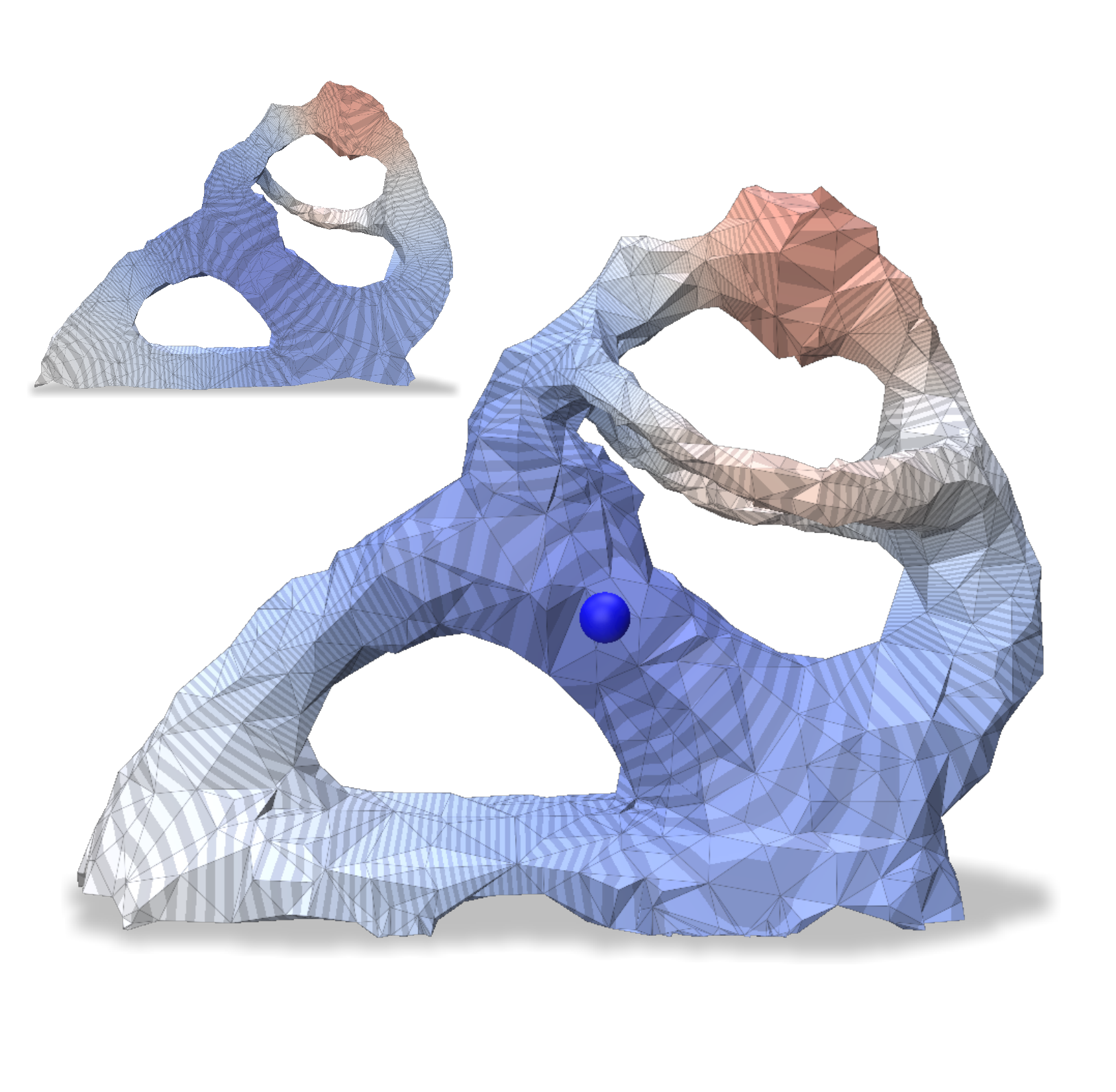} &
			\includegraphics[width=\spdist\columnwidth]{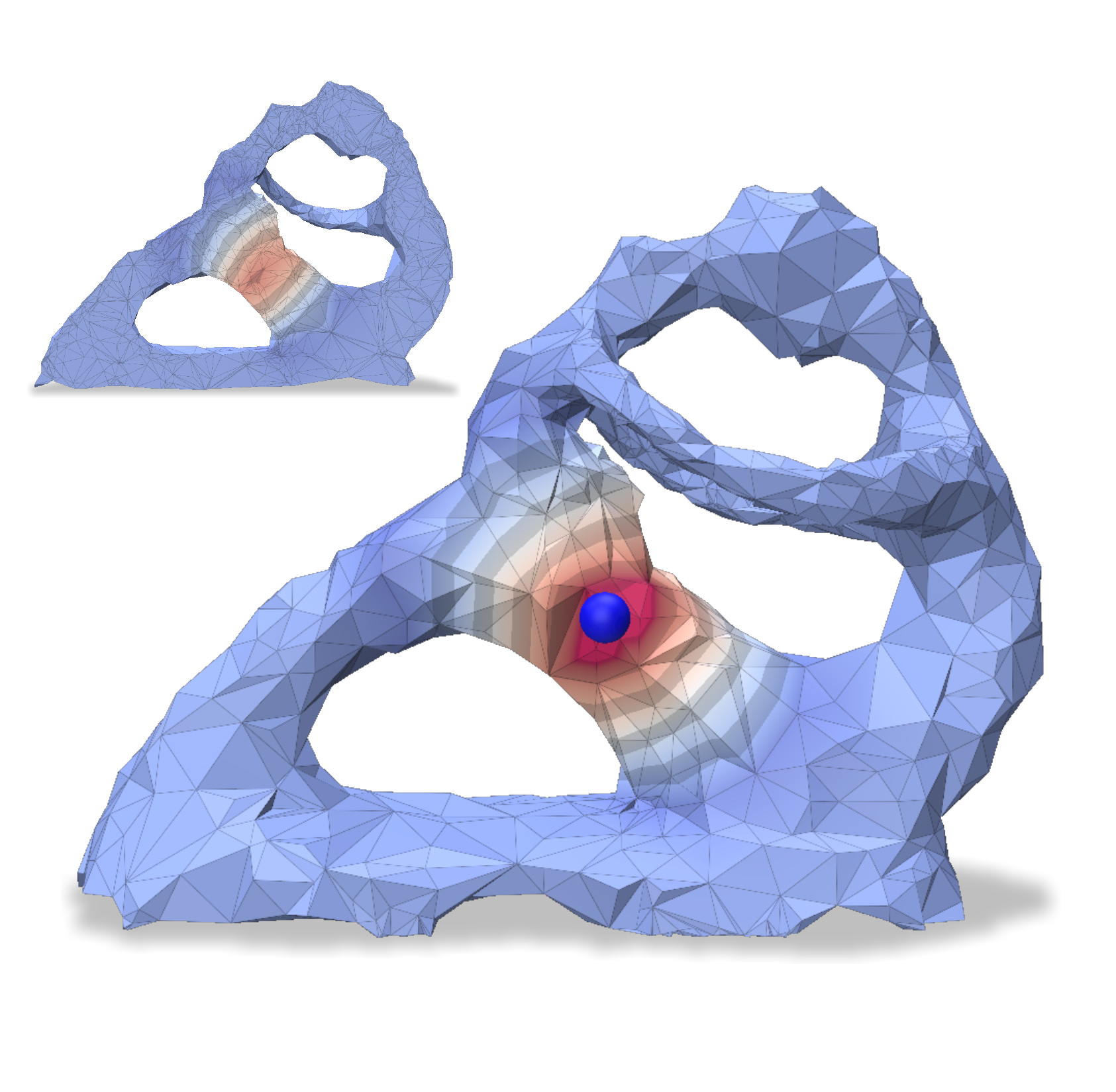} 
			\\
			\rotatebox{90} {Baseline} &
			\includegraphics[width=\monkeysize\textwidth]{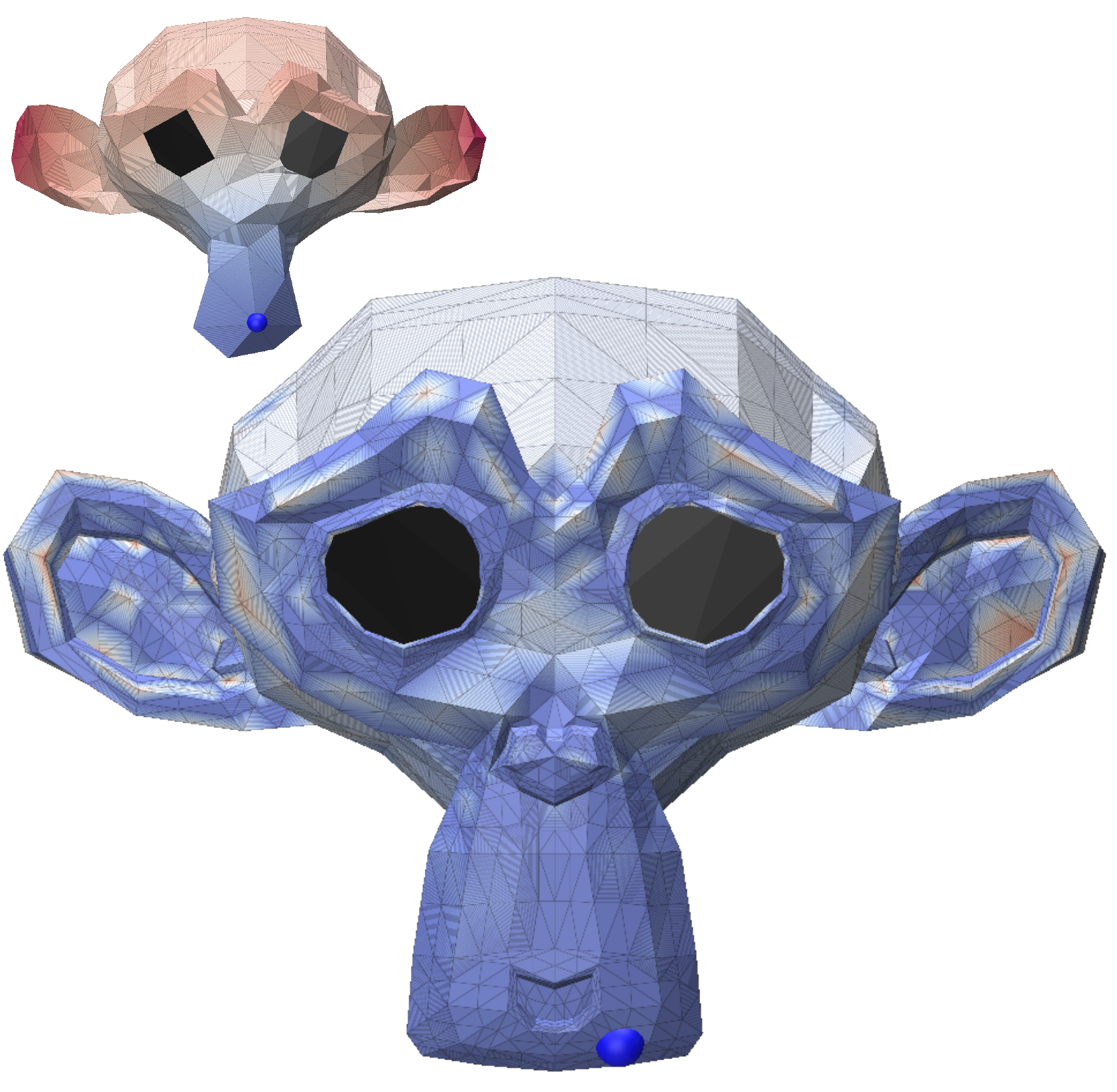} &
			\includegraphics[width=\monkeysize\textwidth]{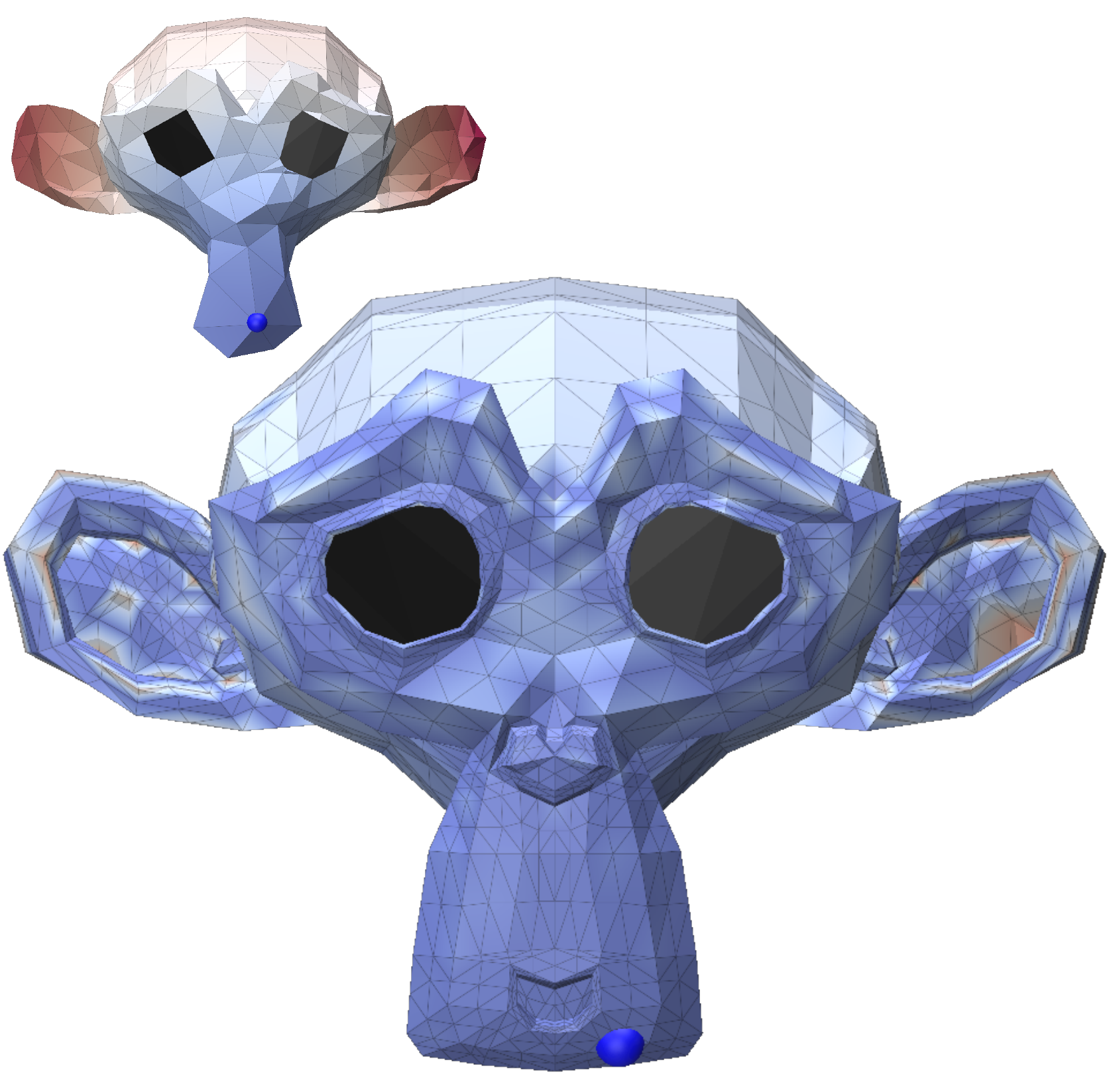} &
			\includegraphics[width=\monkeysize\textwidth]{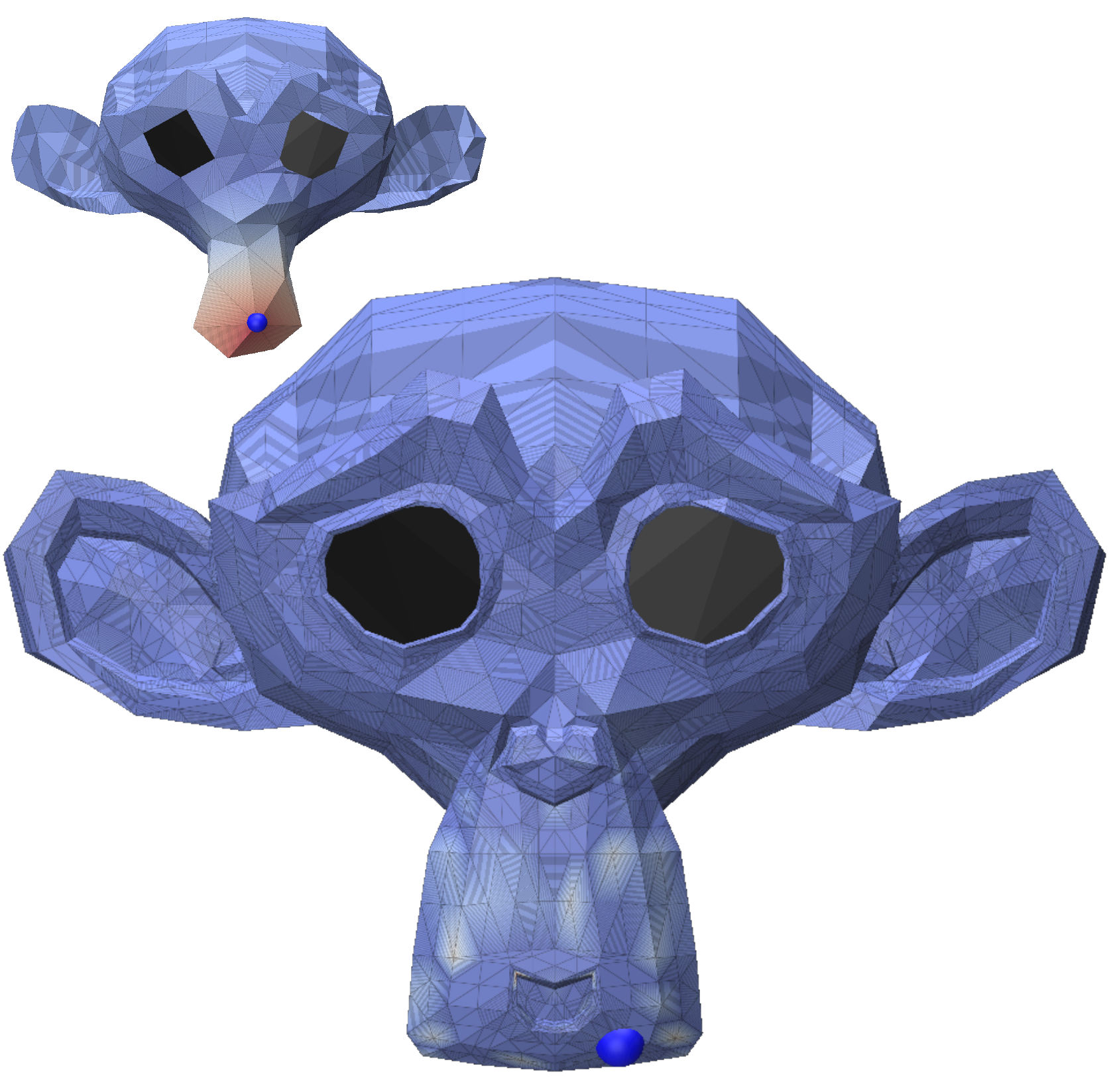} &		
			\rotatebox{90} {Ours - Lifted} &
			\includegraphics[width=\spdist\columnwidth]{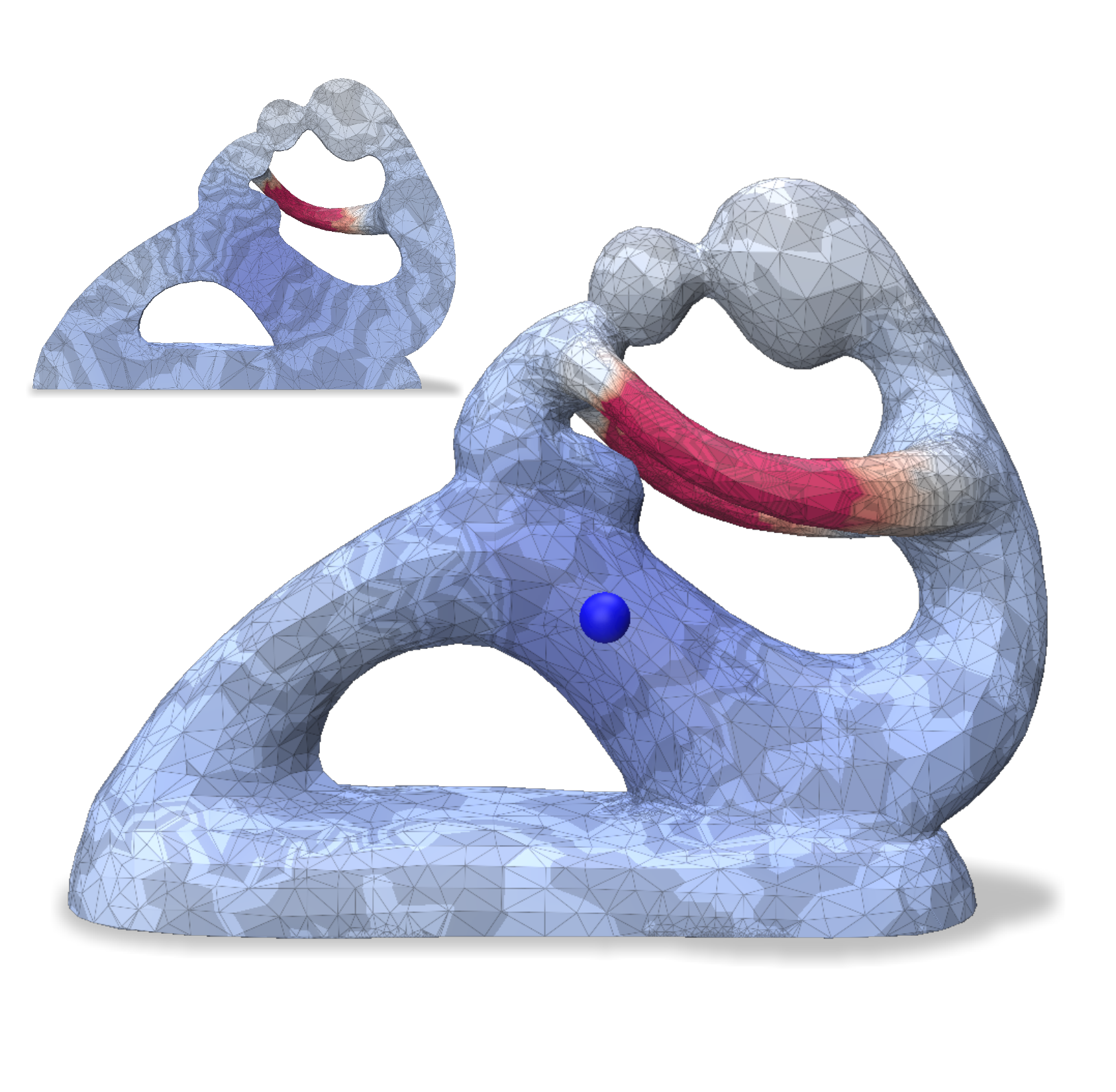} &
			\includegraphics[width=\spdist\columnwidth]{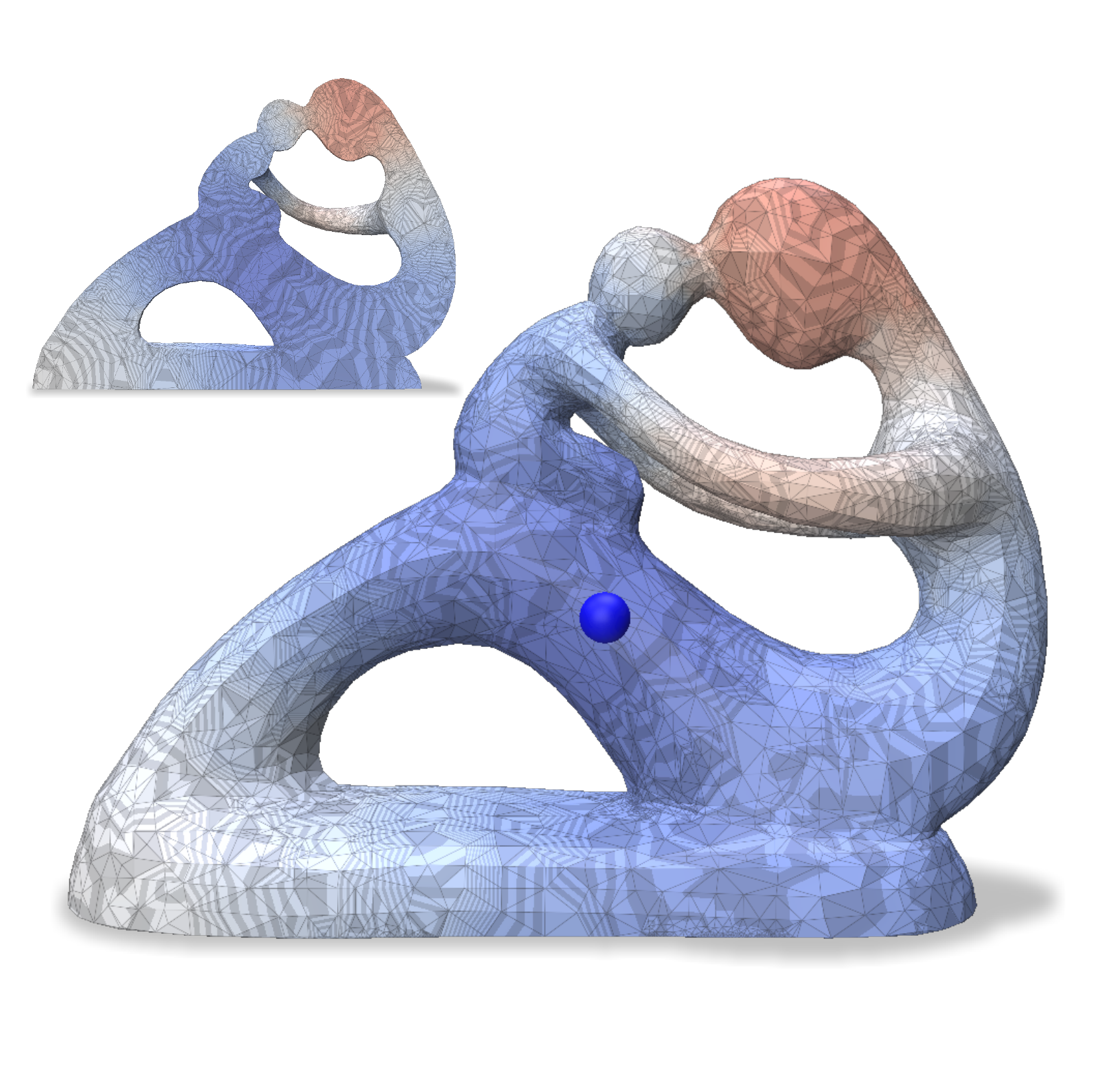} &
			\includegraphics[width=\spdist\columnwidth]{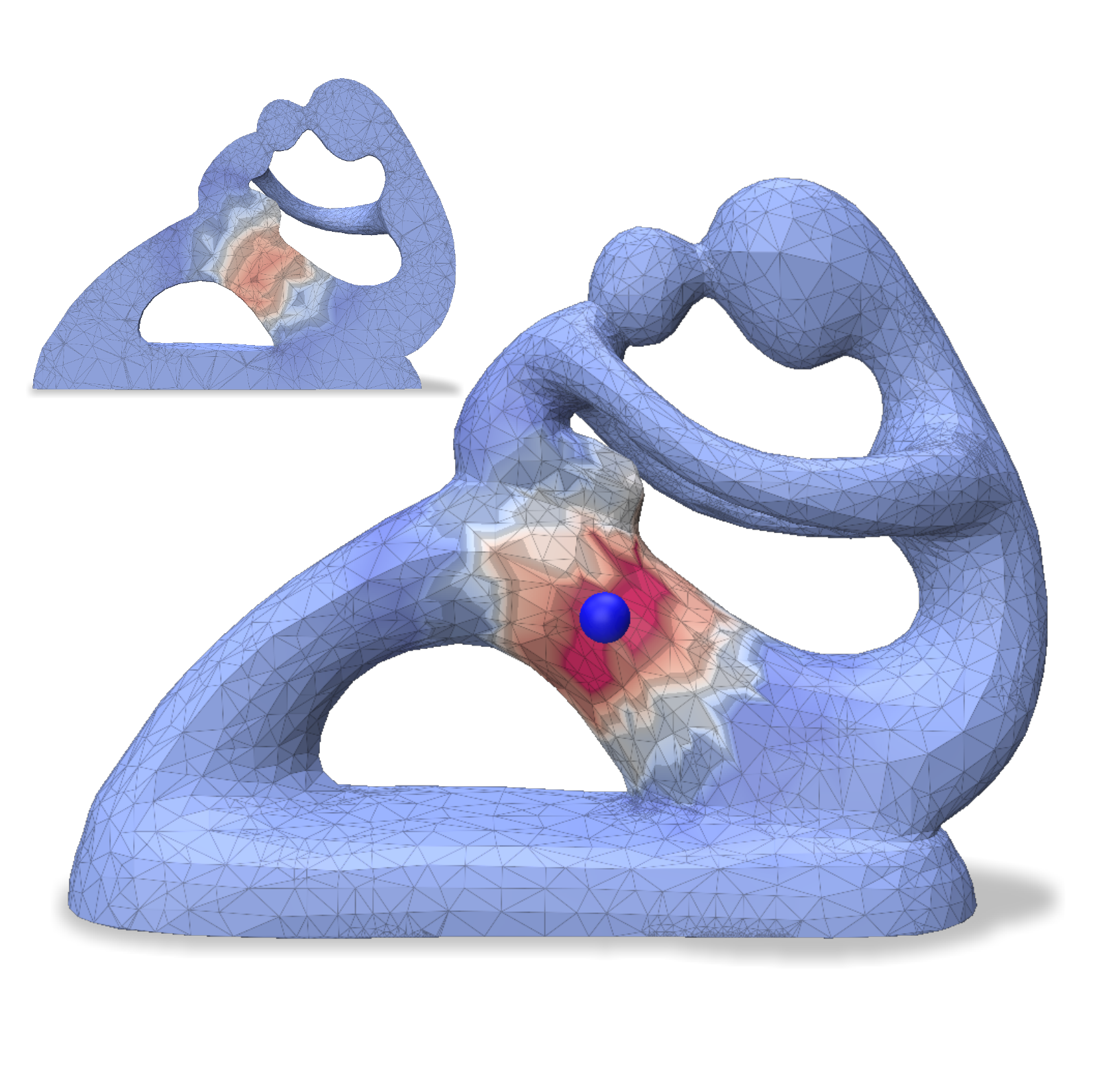} 		 \\
			\multicolumn{4}{c|}{
				\begin{tabular}{@{}l|@{\;\;\;}c@{\;\;\;} >{\columncolor[gray]{0.96}}c@{\;\;\;}c@{\;\;\;} >{\columncolor[gray]{0.96}}c@{\;\;\;}c@{\;\;\;} >{\columncolor[gray]{0.96}}c@{\;\;\;}c@{\;\;\;} >{\columncolor[gray]{0.96}}c@{}} 
					\toprule
					& \small{diff.}  & \small{biharm.} & \small{commute} &   \small{ WKS} & \small{$\text{WKD}$} &            \small{HKS} & $\text{HKD}$ \\
					\midrule
					ours                    &     $\mathbf{0.009}$ & $\mathbf{.00002}$ & $\mathbf{.0009}$  &   $\mathbf{66.06}$ &                    $\mathbf{0.007}$ & $\mathbf{0.003}$ &                    $\mathbf{0.026}$ \\
					baseline     &     0.324 &                 0.033 &                 0.008 & 1292 &                    0.656 & 0.228 &                    0.411 \\
					\bottomrule
				\end{tabular}						
			} &
			\multicolumn{4}{c}{
				\begin{tabular}{@{\;\;\;}c@{\;\;\;} >{\columncolor[gray]{0.96}}c@{\;\;\;}c@{\;\;\;} >{\columncolor[gray]{0.96}}c@{\;\;\;}c@{\;\;\;} >{\columncolor[gray]{0.96}}c@{\;\;\;}c@{\;\;\;} >{\columncolor[gray]{0.96}}c@{}}
					\toprule
					\small{diff.}  & \small{biharm.} & \small{commute} &   \small{ WKS} & \small{$\text{WKD}$} &            \small{HKS} & $\text{HKD}$ \\
					\midrule
					0.699 &      0.739 &   0.019 & 81.6 & 1.597 & .009  &                    2.336 \\
					\bottomrule
				\end{tabular}					
			} 
		\end{tabular}
		
		\caption{\label{fig:dists} Different distance metrics from a point (blue dot on each mesh) are computed on a coarse mesh and visualized using colored isocontours. Our $P^+$ operator enables the ``lifting'' of these distances from the coarse mesh (insets in 2D plots and third row of 3D plots). The tables report mean squared error error for various distance metrics. The insets in the 3D plots depict slices through the mesh.}
	\end{center}
\end{figure*}

\bibliographystyle{ACM-Reference-Format}
\bibliography{hc}



\end{document}